\documentclass[aps,preprint,showpacs,epsfig] {revtex4}
\usepackage{pdfsync}
\usepackage{graphicx}   
\usepackage{epstopdf}
\usepackage{amsmath}

\begin{document}

\title{de Vries behavior of the electroclinic effect in the smectic A$^*$ phase near a biaxiality-induced smectic-A$^*$ -- smectic-C$^*$ tricritical point}

\author{Karl Saunders}

\affiliation{Department of Physics, California Polytechnic State
University, San Luis Obispo, CA 93407, USA}

\email{ksaunder@calpoly.edu}

\date{\today}


\begin{abstract}

Using a generalized Landau theory involving orientational, layering, tilt, and biaxial order parameters we analyze the smectic-A$^*$ and smectic-C$^*$ (Sm-A$^*$ -- Sm-C$^*$) transition, showing that a combination of small orientational order and large layering order leads to Sm-A$^*$ -- Sm-C$^*$ transitions that are either continuous and close to tricriticality or first order. The model predicts that in such systems the increase in birefringence upon entry to the Sm-C$^*$ phase will be especially rapid. It also predicts that the change in layer spacing at the Sm-A$^*$ -- Sm-C$^*$ transition will be proportional to the orientational order. These are two hallmarks of Sm-A$^*$ -- Sm-C$^*$ transitions in de Vries materials. We analyze the electroclinic effect in the Sm-A$^*$ phase and show that as a result of the zero-field Sm-A$^*$ -- Sm-C$^*$ transition being either continuous and close to tricriticality or first order (i.e for systems with a combination of weak orientational order and strong layering order) the electroclinic response of the tilt will be unusually strong. Additionally, we investigate the associated electrically induced change in birefringence and layer spacing, demonstrating de Vries behavior for each, i.e. an unusually large increase in birefringence and an unusually small layer contraction. Both the induced change in birefringence and layer spacing are shown to scale quadratically with the induced tilt angle.

\end{abstract}


\maketitle

\section{Introduction}
\label{Introduction}

\subsection{Background and motivation}
\label{Background and Motivation}

In the last decade there has been significant experimental and theoretical interest in the response of de Vries materials to externally applied electric fields \cite{de Vries review}. In the absence of an applied field de Vries materials exhibit a Sm-A -- Sm-C (or, if chiral, a Sm-A$^*$ -- Sm-C$^*$) transition with an unusually small change in layer spacing and a significant increase in birefringence upon entry to the Sm-C phase. The increase in birefringence is associated with an increase in orientational order. Some de Vries materials exhibit another unusual feature, namely a birefringence that varies nonmonotonically with temperature \cite{Lagerwall, Manna}. Specifically, the birefringence decreases as the Sm-A$^*$ -- Sm-C$^*$  transition is approached from either the low or the high temperature side. de Vries materials generally seem to have unusually small orientational order and follow the phase sequence isotropic -- Sm-A$^*$ -- Sm-C$^*$. In several de Vries materials, the Sm-A$^*$ -- Sm-C$^*$  transition seems to occur close to a tricritical point\cite{HuangDV, Hayashi1, Collings}.

For chiral liquid crystals in general, the application of an electric field to the Sm-A$^*$ phase induces a tilt of the average molecular direction, relative to the layer normal, and hence the optical axis. This phenomenon, known as the electroclinic effect, was first predicted using a symmetry based argument \cite{Meyer} and was then observed experimentally \cite{Garoff and Meyer}. The electroclinic effect led to the development of electro-optic devices using ferroelectric, i.e. chiral liquid crystals. However, the quality of these devices has been limited by the formation of chevron defects, which result from a significant layer contraction associated with the electrically induced molecular tilt.

Ferroelectric de Vries materials have generated considerable excitement because in the Sm-A$^*$ phase they exhibit an unusual electroclinic effect:  a very large reorientation of the optical axis with a very small associated layer contraction. Additionally, there is a very large increase in the birefringence. Aside from being scientifically interesting, such an electroclinic effect makes ferroelectric de Vries materials strong candidates for liquid crystal devices that have large electro-optical response without the associated problem of chevron defects. 

\begin{figure}
\includegraphics[scale=0.8]{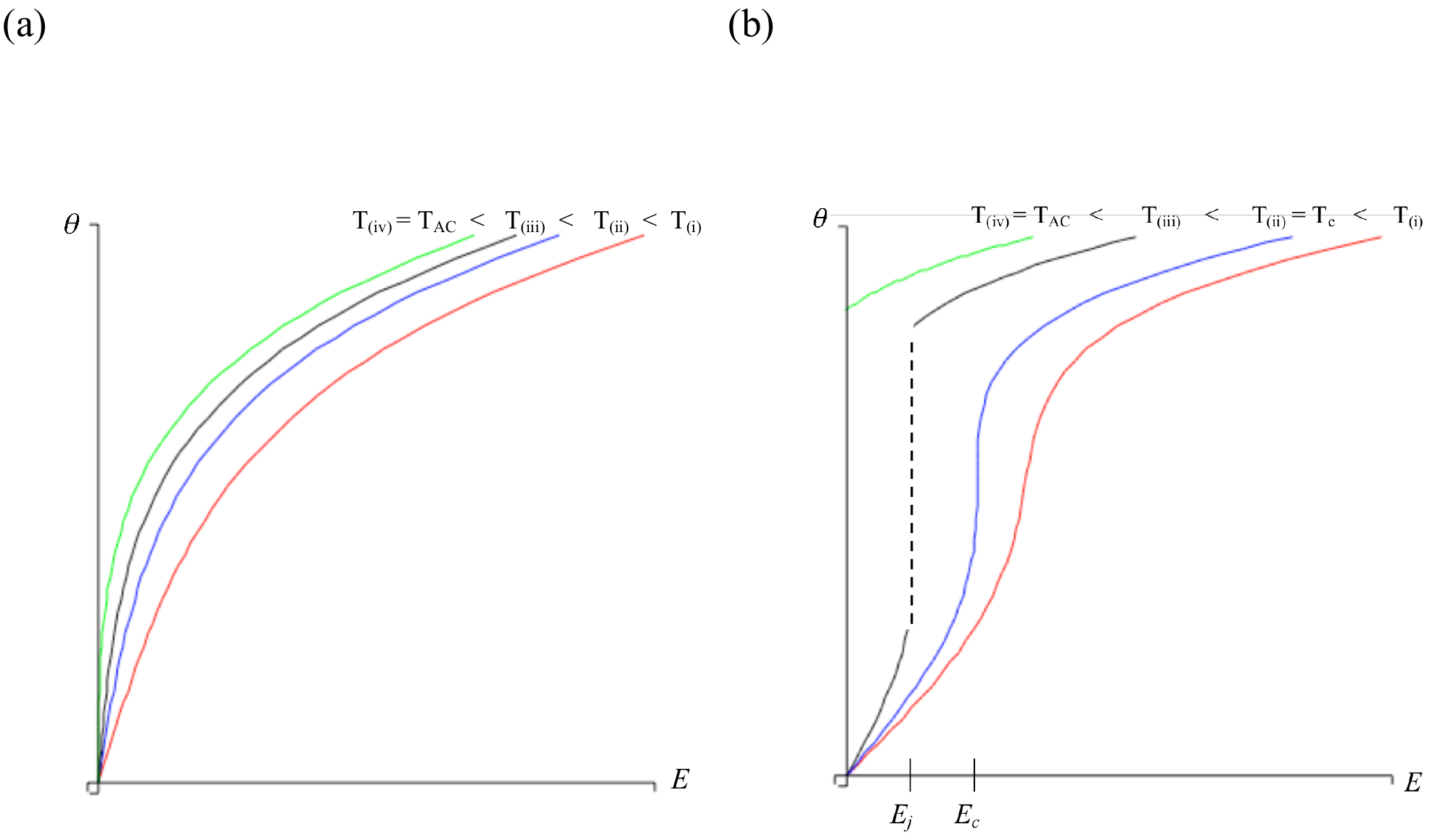}
\caption{A schematic representation of different types of electroclinic response curves.
(a) $\theta(E)$ for materials with continuous Sm-A$^*$ -- Sm-C$^*$ transitions. The curves (i)-(iv) have progressively smaller $T$ with curve (iv) having $T=T_{_{AC}}$. The susceptibility $\chi = \frac{d\theta}{dE}$ is largest at $E=0$, and monotonically decreases as $E$ is increased. The response increases as temperature, $T$, is lowered towards the Sm-A$^*$ -- Sm-C$^*$ transition temperature, $T_{_{AC}}$, with the zero-field susceptibility $\chi_{_0}$ diverging as $T$ approaches $T_{_{AC}}$. (b) $\theta(E)$ for materials with first order Sm-A$^*$ -- Sm-C$^*$ transitions. Curve (i) shows the response for $T>T_c$, a critical temperature. In this case the response is continuous but ``superlinear'', corresponding to positive curvature at small fields followed by negative curvature at large fields. $\chi$ is largest where the curvature changes sign. As $T$ is lowered towards $T_c$ this value of $\chi$ diverges. On curve (ii), corresponding to $T=T_c$, $\chi$ diverges at $E_c$. For $T<T_c$ the response becomes discontinuous, and $\theta$ jumps at field $E_j$. The value of $E_j$ decreases continuously to zero as $T$ is lowered towards $T_{_{AC}}$. Curves (iii) and (iv) correspond to $T_{_{AC}}<T<T_c$ and $T=T_{_{AC}}$ respectively. The value of $\chi_{_0}$ remains finite as $T$ is lowered towards $T_{_{AC}}$.}
\label{response curves}
\end{figure}

There are some details of the electro-optical response in the Sm-A$^*$ phase of de Vries materials that merit further discussion. An important characterization of the electroclinic effect is the response curve $\theta(E)$, where $\theta$ is the tilt of the optical axis and $E$ is the strength of the applied electric field. Different types of electroclinic response curves are shown schematically in Fig.~\ref{response curves} and it can be seen that they are generally nonlinear \cite{continuous response, discontinuous response}.  As shown in Fig.~\ref{response curves}(a), for systems with a continuous Sm-A$^*$ -- Sm-C$^*$ transition $\theta(E)$ is also continuous. As is typical for the electroclinic effect, the curvature $\frac{d^2\theta }{d E^2}<0$ so the susceptibility $\chi = \frac{d\theta}{dE}$ is largest at $E=0$. The zero-field susceptibility $\chi_{_{0}}$ diverges as the temperature $T$ is lowered towards the Sm-A$^*$ -- Sm-C$^*$ transition temperature, $T_{_{AC}}$. For systems with a first order Sm-A$^*$ -- Sm-C$^*$ transition the situation is quite different. For temperatures above a critical temperature $T_c$ the response is continuous but exhibits what has been termed ``superlinear growth". As shown in Fig.~\ref{response curves}(b), this corresponds to positive curvature at small fields followed by negative curvature at large fields. It can also be seen that $\chi$ is largest at the field where the curvature changes sign. As $T$ is reduced towards $T_c$ this value of $\chi$ diverges. For $T<T_c$ the response becomes discontinuous, as shown in Fig.~\ref{response curves}(b), and there is now a jump in the $\theta$ at $E_j$. The value of $E_j$ decreases continuously to zero as $T$ is lowered towards $T_{_{AC}}$. The value of $\chi_{_0}$ remains finite as $T$ is lowered towards $T_{_{AC}}$.

The response of the birefringence, $\Delta n(E)$ in the Sm-A$^*$ phase is also nonlinear and is qualitatively similar to the response of the tilt, $\theta(E)$ \cite{continuous response, discontinuous response}. For systems with a continuous Sm-A$^*$ -- Sm-C$^*$ transition $\Delta n(E)$ is also continuous, while for systems with a first order Sm-A$^*$ -- Sm-C$^*$ transition, $\Delta n(E)$ is continuous with superlinear growth for $T>T_c$ and is discontinuous for $T<T_c$ exhibiting a jump at $E_j$. Remarkably, when $\Delta n(E)$ is plotted parametrically against $\theta^2 (E)$, the scaling is essentially linear, {\it regardless} of the nature (continuous or first order) of the transition \cite{continuous response, discontinuous response}. Equally remarkable is the fact that for a given system, the slope of the linear scaling varies very little with temperature. This means that for any de Vries material the response of the birefringence is well fitted by $\Delta n (E) = \Delta n (0)+k(T) \theta^2 (E)$ where $k(T)$ is a material dependent parameter that has only a very weak temperature dependence. There is less published data on the response of the layer spacing due to the application of an electric field, other than to show that it decreases with increasing field and is unusually small \cite{Spector}.

To date, there have been two theoretical approaches to modeling the unusual electroclinic effect that is displayed by de Vries materials. The first \cite{Collings, Clark, Panov,Ratna} is to use a Langevin model (originally proposed by Fukuda in the context of thresholdless antiferroelectricity \cite{Fukuda}) in conjunction with the assumption of a ``hollow cone'' distribution of the molecular directions. For the sake of brevity we refer to this simply as the hollow cone Langevin model. For a hollow cone distribution, the angle $\beta$ between the long axes of the molecules and the layer normal $\bf \hat N$ has a preferred value $\theta_A$. In the absence of a field the distribution of azimuthal angles, i.e. the projections of the molecular axes onto the layering plane, is uniform, so that the average molecular direction, the director $\bf \hat n$, is parallel to $\bf \hat N$. One motivation for the use of such a distribution is that it would explain the absence of layer contraction at the Sm-C$^*$ transition, because the already tilted molecules need only to align azimuthally in order to reorient $\bf \hat n$ away from $\bf \hat N$ by an angle $\theta_A$. However, it has been pointed out \cite{S Lagerwall} that the hollow cone distribution would have a large negative  value of $S_4$ (corresponding to the $P_4(\cos\beta)$ term in an expansion of the distribution in Legendre polynomials) whereas no Sm-A$^*$ materials have been found with negative values of $S_4$ (de Vries materials seem in general to have very small values of $S_4$). The hollow cone Langevin model yields predictions for the electrical response of the director (via the response of the tilt and azimuthal angles) and the birefringence, but not layer spacing, per se. Rather, it is assumed that the response of the layer spacing will be small, due to the assumption of a hollow cone distribution. 

The hollow cone Langevin model cannot describe systems with response curves of the type shown in Fig.~\ref{response curves}(b), i.e., systems with first order transitions. This has motivated the use of a second type of model, namely that initially presented by Bahr and Heppke in their analysis of a field-induced critical point near the Sm-A$^*$ -- Sm-C$^*$ transition \cite{Bahr Heppke}. While this model provides an accurate description of the response curves, it does not make any predictions regarding the electrical response of the birefringence or layer spacing. Additionally, it does not make any connection to the de Vries behavior of the zero-field Sm-A$^*$ -- Sm-C$^*$ transition.

\subsection{Summary of results}
\label{Summary of Results}

In this article we present and analyze a model that is a chiral extension of the generalized Landau mean field theory that was presented in Refs. \cite{Saunders1, SaundersTCP}. This model is based on an expansion of the free energy density in powers of orientational, layering, tilt and biaxial order parameters. There are chiral couplings of these order parameters to an externally applied field, the effects of which include the electroclinic effect. Our analysis of this chiral model predicts all of the main experimentally observed features of de Vries materials outlined  above: the de Vries behavior (near the zero-field Sm-A$^*$ -- Sm-C$^*$ transition) of layer spacing $d$ and birefringence $\Delta n$, as well as the non-monotonicity of $\Delta n$; proximity of the transition to a tricritical point; the unusually strong electrical response of tilt $\theta(E)$ and birefringence $\Delta n(E)$ in the Sm-A$^*$ phase, along with unusually small layer contraction; the linear scaling of $\Delta n(E)$ vs $\theta^2(E)$, regardless of the nature of the zero-field Sm-A$^*$ -- Sm-C$^*$ transition. Furthermore, {\it all} of these features can be accounted for if the system possesses unusually small orientational order and strong layering order, a combination thought prevalent among de Vries materials. These results do not rely on any particular assumptions about the distribution of the molecular directions, other than that the distribution corresponds to small orientational order.

\subsubsection{Zero-field Sm-A$^*$ -- Sm-C$^*$ transition}
\label{zero-field Sm-A* - Sm-C* Transition}

\begin{figure}
\includegraphics[scale=1]{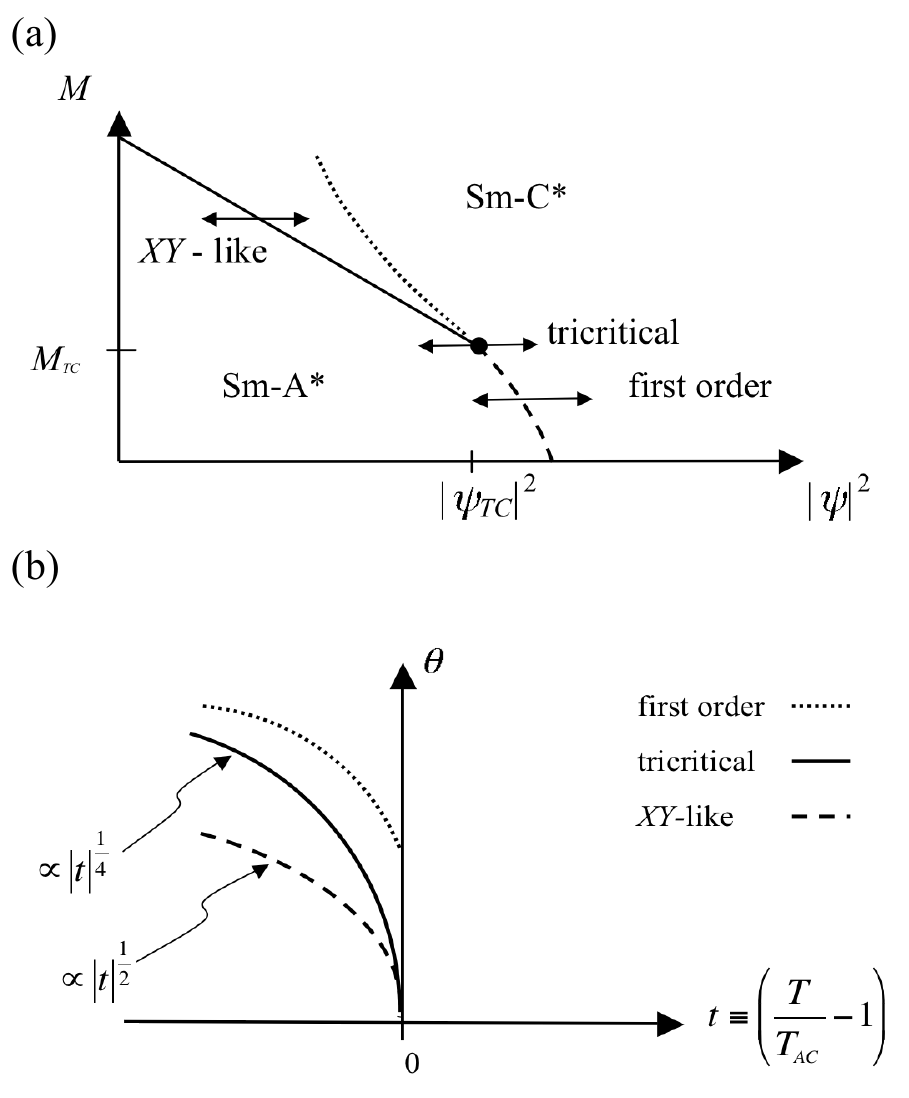}
\caption{(a)The phase diagram in $|\psi|^2$-$M$ space near the tricritical point ($|\psi_{_{TC}}|^2$, $M_{_{TC}}$). The quantity $M$ is a measure of how much orientational order the system possesses and for de Vries materials is effectively athermal. The quantity $|\psi|$ is a measure of the strength of the layering. It is a monotonically decreasing function of temperature so that for a given material, decreasing the temperature corresponds to moving horizontally from left to right. The solid line represents the continuous Sm-A$^*$ -- Sm-C$^*$ boundary while the dashed line represents the first order Sm-A$^*$ -- Sm-C$^*$ boundary. These two boundaries meet at the tricritical point ($|\psi_{_{TC}}|^2$, $M_{_{TC}}$). The dotted line indicates the region in which the behavior crosses over from $XY$-like to tricritical. The region in which the behavior is $XY$-like shrinks to zero as the tricritical point is approached. At the tricritical point the slopes of the first order and continuous Sm-A$^*$ -- Sm-C$^*$  boundaries are equal but the curvatures are not. Also shown, as double ended arrows, are the three distinct classes of transitions: $XY$-like, tricritical and first order. (b) The tilt angle $\theta$ as a function of reduced temperature $t \equiv \left(1-\frac{T}{T_{_{AC}}}\right)$ near the Sm-A$^*$ -- Sm-C$^*$ transition temperature $T_{_{AC}}$, i.e., for $|t| \ll1$. Upon entry to the Sm-C$^*$ phase the growth of the tilt angle scales like $\left| t \right|^{\frac{1}{2}}$ for a mean field $XY$-like transition. For a tricritical transition it scales like $\left| t \right|^{\frac{1}{4}}$ and is thus more rapid. For a first order transition there is a jump in the tilt angle upon entry to the Sm-C$^*$ phase.}
\label{phase diagram}
\end{figure}

Figure \ref{phase diagram} shows the Sm-A$^*$ -- Sm-C$^*$ phase boundary in $|\psi|^2$-$M$ space, where $|\psi|$ and $M$ are the magnitudes of the layering and orientational order parameters, respectively. They will be defined more rigorously in Section \ref{Free energy density for a nonchiral system}. It has been observed that the orientational order in de Vries systems has only a very weak temperature dependence. Along with the fact that the nematic phase does not occur for all known de Vries materials, this implies \cite{Saunders1, SaundersTCP} that the transition to the Sm-C$^*$ phase is driven by an increase in the layering as the temperature decreases. Thus, in the phase diagram of Fig.~\ref{phase diagram}(a), varying the temperature corresponds to a horizontal path. It is important to note that the negative slope of the Sm-A$^*$ -- Sm-C$^*$ phase boundary implies that the smaller the value of $M$, the larger the value of $|\psi|$ at which the Sm-A$^*$ -- Sm-C$^*$ transition occurs. This is consistent with the observation  \cite{de Vries review, S Lagerwall} that de Vries smectics generally have such unusually weak orientational order that their stabilization requires strong layering order, perhaps via micro-segregation.

The zero-field model predicts that a Sm-A$^*$ -- Sm-C$^*$ tricritical point results due to a coupling between biaxiality and tilt. The effect of biaxiality is stronger in systems with small $M$ and large $|\psi|$, so that a tricritical point and associated neighboring first order transition can be accessed by systems with sufficiently small orientational order, $M\leq M_{_{TC}}$. Here $M_{_{TC}}$ is the value of the orientational order at which the system exhibits a tricritical Sm-A$^*$ -- Sm-C$^*$ transition. This is shown in the phase diagram of Fig. {\ref{phase diagram}. 

As usual, for systems with continuous Sm-A$^*$ -- Sm-C$^*$ transitions, the growth of $\theta$ upon entry to the Sm-C$^*$ phase scales like $\theta \propto |t|^\beta$ where, $t=\frac{T}{T_{_{AC}}}-1$, is the reduced temperature and $T_{_{AC}}$ is the Sm-A$^*$ -- Sm-C$^*$ transition temperature \cite{mean field footnote}. Away from the tricritical point the scaling is $XY$-like, so $\beta=0.5$, and at the tricritical point $\beta=0.25$ implying a more rapid growth of $\theta$ at tricriticality, as shown in Fig. {\ref{phase diagram}(b). In the Sm-C$^*$ phase, for $M>M_{_{TC}}$, there is a crossover in the scaling from $XY$-like to tricritical at some reduced temperature $t_*(M)$. As $M$ is lowered towards $M_{_{TC}}$ this crossover $t_*$ shrinks to zero. For $M_0\leq M_{_{TC}}$ the transition Sm-A$^*$ -- Sm-C$^*$ is first order and there is a discontinuous jump in $\theta$ at the transition, also shown in Fig. {\ref{phase diagram}(b). 

The behavior of the birefringence near the zero-field Sm-A$^*$ -- Sm-C$^*$ transition is essentially the same as that for the Sm-A -Sm-C transition. This behavior is best described in terms of the fractional change in birefringence $\Delta_{n} \equiv \frac{\Delta n -\Delta n_{_{AC}}}{\Delta n_{_{AC}}} $, where $\Delta n _{_{AC}}$ is the birefringence in the Sm-A$^*$ phase right at the Sm-A$^*$ -- Sm-C$^*$ boundary.  As discussed in Ref.~\cite{SaundersTCP} we find that upon entry to the Sm-C$^*$ phase, for any of the three types of transitions ($XY$-like, tricritical, first order), $\Delta_{n} $ of a de Vries type material will grow according to $\Delta_{n} \propto \theta^2$. While the dependence of $\Delta_{n}$ on $\theta$ is the same for all three types of transitions, its dependence on temperature is not the same because, as shown in the Fig.~\ref{phase diagram}(b),  $\theta$ scales differently with temperature for each type of transition. For an $XY$-like transition the growth of $\Delta_{n}$ will be linear, $\propto \left| t \right|$, while for a transition at tricriticality it scales like $ \left| t \right|^{\frac{1}{2}}$ and is thus more rapid. For a first order transition there will be a jump in the tilt angle and, therefore, an associated jump in $\Delta_{n}$, although near tricriticality, where the transition is only weakly first order, the jump will be small. Thus, the rapid growth of birefringence observed in de Vries materials can be attributed to the proximity of the system's Sm-A$^*$ -- Sm-C$^*$ transition to a tricritical point, which as discussed above, can in turn be attributed to unusually small orientational order. Additionally, we predict the possibility of a weakly temperature dependent birefringence that {\it decreases} as the zero-field Sm-A$^*$ -- Sm-C$^*$ transition is approached from the Sm-A$^*$ phase, which as discussed above, is an unusual feature that has been observed experimentally \cite{Lagerwall, Manna}. 

Similarly, the behavior of the layer spacing $d$ near the zero-field Sm-A$^*$ -- Sm-C$^*$ transition is essentially the same as that for the Sm-A -Sm-C transition, and is best described in terms of the layer contraction $\Delta_d \equiv (d_{AC}-d_C)/d_{AC}$, where $d_{AC}$ and $d_C$ are the layer spacing in the Sm-A$^*$ phase (right at the Sm-A$^*$ -- Sm-C$^*$ boundary) and in the Sm-C$^*$ phase, respectively. We find that for any of the three possible types of transitions, $\Delta_d \propto M_0 \theta^2$. Thus, for unusually small orientational order $M_0$, the layer contraction is unusually small, and therefore de Vries-like.

\subsubsection{Electroclinic Effect in the Sm-A$^*$ Phase}
\label{Electroclinic Effect in the Sm-A* Phase}

With the application of an electric field of strength $E$, we show that our generalized Landau model predicts the following relationship between the induced tilt, $\theta$, and $E$:
\begin{eqnarray}
E= \alpha_e(t,M,d) \theta + \beta_e (M,d) \theta^3 + \gamma_e (M,d) \theta^5\;.
\label{E(c)Intro}
\end{eqnarray}
This relationship is completely analogous to that presented by Bahr-Heppke in the context of a field induced critical point near the Sm-A$^*$ -- Sm-C$^*$ transition \cite{Bahr Heppke}. However, our derivation of Eq.~(\ref{E(c)Intro}) from the more basic level of a generalized Landau theory (in terms of layering and orientational order parameters) allows us to relate the coefficients $\alpha_E(t,M,d)$, $\beta_E(M,d)$, $\gamma_E(M,d)$ to the orientational order, $M$, and the layer spacing, $d$, in the system. This allows us to do two important things. Firstly, we can determine the nature of the response $\theta(E)$ (i.e., continuous with decreasing slope, superlinear or discontinuous) based on the degree of orientational order $M$ in the system. Secondly, using Eq.~(\ref{E(c)Intro}) along with the rest of the generalized free energy, we can determine the electrical response of the birefringence (which is proportional to the $M$) and the layer spacing $d$.

The nature of the response depends crucially on the sign of $\beta_e(M,d)$. We find that $\beta_e(M,d)\propto (M-M_{_{TC}})$. Thus, for sufficiently large orientational order $M\geq M_{_{TC}}$, i.e. for systems with a continuous Sm-A$^*$ -- Sm-C$^*$ transition, $\beta_e>0$ and the response is continuous with susceptibility decreasing as $E$ is increased. The response at the continuous Sm-A$^*$ -- Sm-C$^*$ transition for small fields scales like $\theta \propto E^{\frac{1}{\delta}}$. Away from tricriticality ($M>M_{_{TC}}$) $\delta=3$ while at tricriticality ($M=M_{_{TC}}$) $\delta=5$ and the response is significantly stronger. For sufficiently small orientational order $M\leq M_{_{TC}}$, i.e. for systems with a first order Sm-A$^*$ -- Sm-C$^*$ transition, $\beta_e>0$. In this case for sufficiently large temperature $T>T_c$ the response is superlinear while for $T<T_c$ the response curve $\theta(E)$ becomes ${\cal S}$ shaped and there is a jump in $\theta$ as the field is increased through $E_j$. At $T=T_c$ the susceptibility diverges at $E_c$, and, as shown by Bahr and Heppke, the corresponding point $(T_c, E_c, \theta(T_c,E_c))$ is a critical point. Thus, like the rapid growth of the zero-field birefringence at the Sm-A$^*$ -Sm -C$^*$ transition, the strong electrical response of the tilt in de Vries materials can be attributed to the proximity of the system's Sm-A$^*$ -- Sm-C$^*$ transition to a tricritical point. This can in turn be attributed to the unusually small orientational order of de Vries materials.

In describing the change in birefringence due to an applied field, it is useful to define the fractional change of the birefringence due to the applied electric field, $\Delta_{n}(E)\equiv \frac{\Delta n(E) -\Delta n(0)}{\Delta n(0)}$ where $\Delta n(E)$ is the birefringence in the presence of a field of magnitude $E$. We show that regardless of the nature of the transition (and hence the response) $\Delta_{n}(E)$ scales linearly with $\theta^2(E)$  i.e.,
\begin{eqnarray}
\Delta_{n}(E) = \eta(T) \theta^2 (E)\;.
\label{DeltanE}
\end{eqnarray}
This scaling, shown in Fig.~\ref{Delta_d_E&Delta_n_E}(a), is consistent with experiment \cite{continuous response,discontinuous response}. The dimensionless constant $\eta(T)\propto |\psi(T)|^2/d(T)^2$ depends on temperature via its dependence on layering strength $|\psi(T)|$ and layer spacing $d(T)$. Since both $\frac{d}{dT}|\psi(T)|$ and $\frac{d}{dT}d(T)$ have the same sign (i.e., negative), it is possible that $\eta(T)$ is only weakly dependent on temperature, which would be consistent with experiment. The relationship given in Eq.~(\ref{DeltanE}) means that an unusually strong, e.g. discontinuous, electrical response of the tilt will imply an unusually strong response, e.g. discontinuous,  of the birefringence, which again is consistent with experiment.  
\begin{figure}
\includegraphics[scale=1]{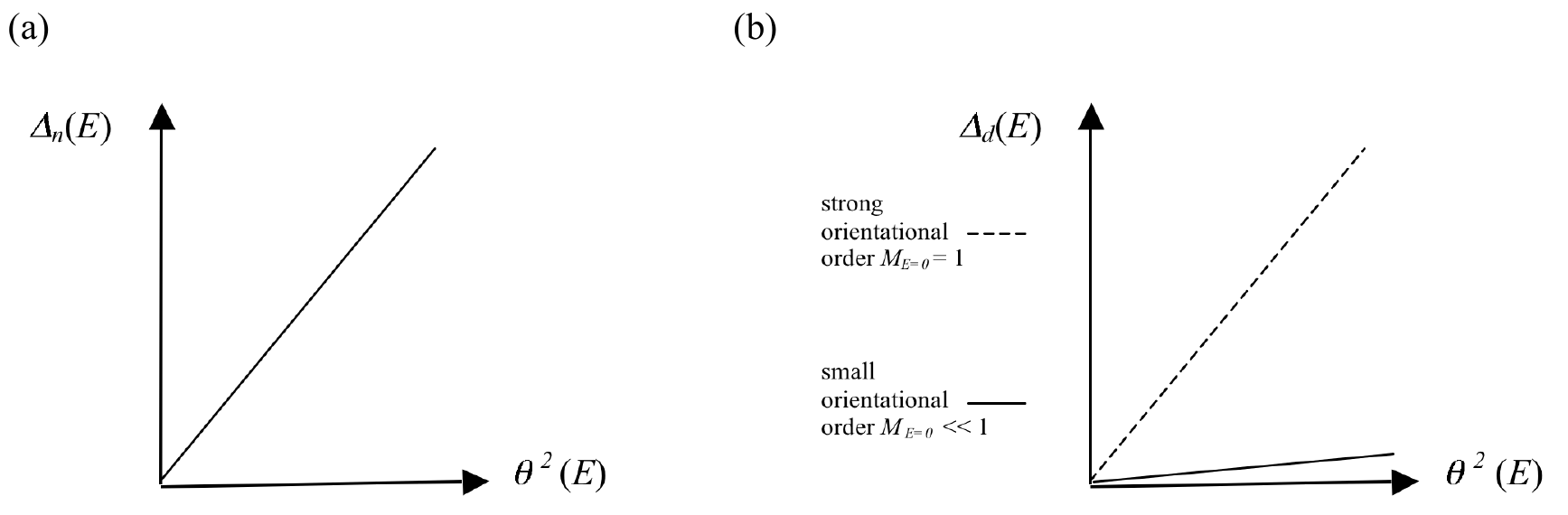}
\caption{(a)A plot of the fractional change of the birefringence due to applied electric field, $\Delta_{n}(E)\equiv \frac{\Delta n(E) -\Delta n(0)}{\Delta n(0)}$ versus the square of the induced tilt, $\theta^2(E)$. For any type of transition (and hence any type of response of $\theta(E)$) we find that the scaling of $\Delta_{n}(E)$ with $\theta^2(E)$ is linear. The model predicts the possibility of a weakly temperature dependent slope $\eta(T)$. (b) A plot of the layer contraction due to applied electric field,  $\Delta_{d} (E)\equiv\frac{d(E) -d(0)}{d(0)}$ versus the square of the induced tilt, $\theta^2(E)$. For any type of transition (and hence any type of response of $\theta(E)$) we find that the scaling of $\Delta_{d} (E)$ with $\theta^2(E)$ is linear. The slope of the scaling is  proportional to $M_{E=0}$, the value of the zero-field orientational order, which for de Vries materials is unusually small. Two plots are shown, one for a system with small orientational order $M_{E=0} \ll 1$, for which the contraction will be small, and one for a system with strong orientational order $M_{E=0}  \approx 1$, for which the contraction will be sizable.}
\label{Delta_d_E&Delta_n_E}
\end{figure}

Similarly, the layer spacing $d(E)$ is affected by the field, and the layer contraction $\Delta_{d}(E)\equiv\frac{d(E) -d(0)}{d(0)}$ also scales linearly with $\theta^2(E)$ regardless of the nature of the transition (and hence the response), i.e.,
\begin{eqnarray}
\Delta_{d} (E) \propto M_{E=0} \theta^2 (E)\;,
\label{Delta d(E)}
\end{eqnarray}
where $M_{E=0}$ is the value of the zero-field orientational order, which for de Vries materials is unusually small. Thus, for de Vries materials the contraction of the layers associated with the electroclinic effect will also be unusually small. As with the birefringence, the shape of response curve $d(E)$ will be nonlinear and discontinuous if $\theta(E)$ is. However, regardless of the shape, if $M_{E=0}$ is small the layer contraction will be too. This is summarized in in Fig.~\ref{Delta_d_E&Delta_n_E}(b). As discussed above, there is less published data on the response of the layer spacing other than to show that it is small. Further experimental investigation of the response could be in interesting, in order to see if it is consistent with Eq.~(\ref{Delta d(E)}) above.

\subsection{Outline}
\label{Outline}

The remainder of this article is organized as follows. In Section \ref{Nonchiral Model} we review the nonchiral generalized Landau theory. This is done with a view to using it as the basis of our chiral model and we focus in particular on the parts of model that are important for the analysis of the electroclinic effect. Additionally, we review the results for the nonchiral zero-field phase diagram, as it will be argued later that the phase diagram for a chiral system is is essentially the same. In Section \ref{Incorporating the effects of chirality and external fields to the free energy density} we generalize the model to reflect the presence of chirality and an external field. The general approach to doing so is to add the relevant chiral terms and field dependent terms. To strike a balance between making the model realistic and making it manageable, we are selective in what we add to reflect the presence of chirality and a field. The justification behind our selection is discussed in Section \ref{Incorporating the effects of chirality and external fields to the free energy density}. In Section \ref{Response of Tilt} we analyze the response of the tilt to a field applied to the Sm-A$^*$ phase. In Section \ref{Response of Birefringence and Layer Spacing} we analyze the response of the birefringence and layer spacing to a field applied to the Sm-A$^*$ phase. We provide a brief recap of our results in Section \ref{Summary}. The Appendix includes details of the analysis from Section \ref{Response of Birefringence and Layer Spacing}.

\section{Model and results for a nonchiral system}
\label{Nonchiral Model}

In constructing the free energy density for a chiral smectic we follow the usual strategy of starting with a nonchiral free energy density and then adding the terms that reflect the breaking of the chiral symmetry and the presence of a field. In this section we discuss the nonchiral model and results.

\subsection{Free energy density for a nonchiral system}
\label{Free energy density for a nonchiral system}

The nonchiral free energy density includes orientational, tilt (azimuthal), biaxial and layering order parameters. The complex layering order parameter $\psi$ is defined via the density $\rho=\rho_0+$ Re$(\psi e^{i \bf q \cdot r})$ with $\rho_0$ constant and ${\bf q}$ the layering wavevector, the arbitrary direction of which is taken to be $z$.  The remaining order parameters are embodied in the usual second rank tensor orientational order parameter $\cal Q$, which is most conveniently expressed as
\begin{eqnarray}
Q_{ij}&=& M [ (-\cos(\alpha)+\sqrt{3}\sin(\alpha))e_{1i} e_{1j} \nonumber\\ 
&+&(-\cos(\alpha)-\sqrt{3}\sin(\alpha))e_{2i} e_{2j}+2\cos(\alpha)e_{3i} e_{3j}] \;,
\label{Q}
\end{eqnarray}
where ${\bf \hat e_3} = {\bf c} + \sqrt{1-c^2}{\bf \hat z}$ is the average direction of the molecules' long axes, (i.e., the director). Here, in either the Sm-A or Sm-C phase, ${\bf \hat z}$ is normal to the plane of the layers. The projection, ${\bf c}$, of the director onto the layers is the order parameter for the Sm-C phase. The other two principal axes of ${\cal Q}$ are given by ${\bf \hat e_1} = {\bf \hat z} \times {\bf \hat c}$ and ${\bf \hat e_2} = \sqrt{1-c^2}{\bf \hat c} - c {\bf \hat z}$. These unit eigenvectors are shown in Fig.~\ref{Eigenvectors}. The amount of orientational order is given by $M\propto \sqrt{ Tr({\cal Q}^2)}$, which is proportional to the birefringence. The degree of biaxiality is described by the parameter $\alpha$. The Sm-A phase is untilted (${\bf c} = {\bf 0}$) and uniaxial ($\alpha=0$), while the Sm-C phase is tilted (${\bf c} \neq {\bf 0}$) and biaxial ($\alpha\neq 0$). From Fig.~\ref{Eigenvectors} it can be seen that the angle $\theta$, by which the optical axis tilts, can be related to $c$ via $c=\sin(\theta)$.
\begin{figure}
\includegraphics[scale=1.4]{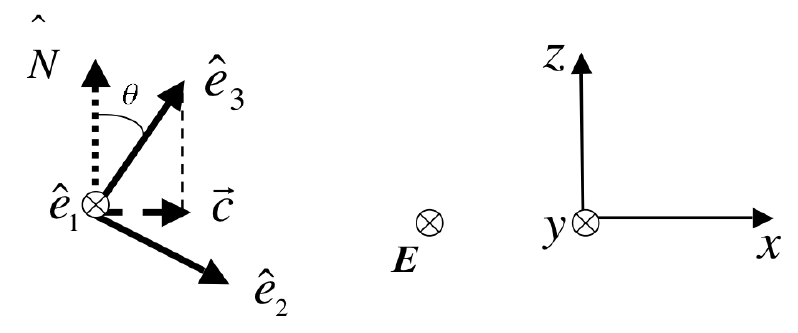}
\caption{The unit eigenvectors, ${\bf \hat e_1}$, ${\bf \hat e_2}$, ${\bf \hat e_3}$ of the orientational order tensor ${\cal Q}$. These are shown as solid arrows, with ${\bf \hat e_1}$ pointing into the page. Also shown, as a dotted arrow, is the layering direction ${\bf \hat N}$, which is normal to the plane of the layers. We choose this as our ${\bf \hat z}$ direction. The eigenvector ${\bf \hat e_3}$ corresponds to the average direction of the molecules' long axes. The order parameter, ${\bf c}$, for the $C$ phase is the projection of ${\bf \hat e_3}$ onto the plane of the layers, and is shown as a dashed arrow. The angle $\theta$, by which the optical axis tilts, is also shown. This is the arrangement that corresponds to the lowest energy state if the applied electric field points into the page. Taking this direction to be ${\bf \hat y}$, i.e., ${\bf E}=E{\bf \hat y}$ implies that  ${\bf c}$ points in the ${\bf \hat x}$ direction. }
\label{Eigenvectors}
\end{figure}

The nonchiral generalized free energy was presented previously \cite{SaundersTCP} as a sum, $f= f_{Q}+f_{\psi}+f_{Q \psi}$, of orientational ($f_{Q}$), layering ($f_\psi$), and coupling ($f_{Q \psi}$) terms. The orientational term consists of terms $\propto (Tr({\cal Q}^n)$, with integer $n>1$. The layering term consists of terms $\propto q^{2n}|\psi|^{2m}$ with integers $n\geq 0$, $m > 0$. The coupling term $f_{Q \psi} $ consists of real, scalar combinations of ${\bf q}$, ${\cal Q}$ and $\psi$, e.g., $q_i q_j Q_{ij}  |\psi|^2$. To make the analysis tractable, the coefficients of these coupling terms were (and will be) assumed to be small. Minimization with respect to the biaxiality $\alpha$ yielded the nonchiral free energy density $f\approx f_M + f_{\psi} + f_{M\psi} + f_c$. The pieces $f_M$ and $f_{\psi}$ only involve the orientational and layering order parameter $M$ and $\psi$ respectively, and are given by
\begin{eqnarray}
f_{M} = \frac{1}{2} r_n M^2 - \frac{1}{3} w M^3 + \frac{1}{4} u_n M^4
 \;,
\label{H_M}
\end{eqnarray}
and
\begin{eqnarray}
f_{\psi} = \frac{1}{2} r_s |\psi|^2 + \frac{1}{4} u_s |\psi|^4
+\frac{1}{2} K (q^2-q_0^2)^2|\psi|^2 \;.
\label{f_Psi}
\end{eqnarray}
The coefficients $w$, $u_n$, $u_s$, and $K$ are positive. As usual in Landau theory, the parameters $r_n$ and $r_s$ are monotonically increasing functions of temperature and control the ``bare'' orientational and layering order parameters, $M_0$ and $\psi_0$ respectively. By ``bare'' we mean the values the order parameters would take on in the absence of the coupling term $f_{Q \psi}$ and an externally applied field. Similarly, the constant $q_0$ is the bare value of the layering wavevector. From Eq.~(\ref{H_M}) and Eq.~(\ref{f_Psi}) above, we immediately find $M_0(r_n)= (w+\sqrt{w^2-4u_n r_n})/2u_n$ and $|\psi_0|=\sqrt{-r_s/u_s}$. As discussed in the Introduction, de Vries behavior occurs in materials where the layering and orientational order parameters are the primary and secondary order parameters respectively. This would imply a virtually athermal $r_n$ (and thus, an athermal $M_0$), so that for a given material $M_0$ can be thought of as a fixed quantity. This means that the temperature variation in orientational order $M$ is effectively due to its coupling to the temperature dependent layering, i.e. via $f_{M\psi}$ and $f_c$. The term $f_{M\psi}$ is given by
\begin{eqnarray}
f_{M\psi} &=& q^2  |\psi|^2 M \left(-a(q^2)+b|\psi|^2+ 2g M + h q^2 M \right)
 \;,
\label{H_M psi}
\end{eqnarray}
where $a(q^2)=a_0 +a_1(q^2-q_0^2)$. The coefficients $a_0$, $a_1$, $b$, $g$ and $h$ are positive and, as discussed above, are treated perturbatively throughout \cite{Notation Change}. For notational simplicity, we suppress the explicit $q$ dependence of $a$, i.e. we use $a=a(q^2)$. The coupling term $f_c$ involves the tilt (azimuthal) order parameter ${\bf c}$ and is given by
\begin{eqnarray}
f_{c} = \frac{1}{2} r_c c^2 + \frac{1}{4} u_c c^4 +\frac{1}{6} v_c c^6 \;.
\label{f_c}
\end{eqnarray}
The coefficients  $r_c$, $u_c$, $v_c$ are given by $r_c =3 a q^2 |\psi|^2 M \tau$, $u_c = 9 \mu h q^4 |\psi|^2 M^2$ and $v_c = \frac{81}{4} s q^6 |\psi|^2 M^3$, with $s$ another coupling coefficient that is treated perturbatively throughout. The parameter $\tau = 1-\frac{b |\psi|^2 + (g+2hq^2)M}{a}$ controls the zero-field transition. The proximity of the zero-field transition to tricriticality is measured by the tricritical proximity parameter $\mu$ which will be discussed below. 

\subsection{Zero-field Sm-A -- Sm-C transitions}

\subsubsection{Continuous Sm-A -- Sm-C transition}

At the continuous Sm-A -- Sm-C transition the dimensionless parameter $\tau$ and thus $r_c$ changes sign. For materials, such as de Vries smectics, with orientational order $M$ that is weakly temperature dependent, this transition occurs due to the layering order $|\psi|$ increasing as temperature decreases. Using the above expression for $\tau$, the continuous transition temperature $T_0$ is defined via $|\psi_0(T_0)|^2 = (a_0-(g+2hq_0^2)M_0)/b$. Figure \ref{detailed phase diagram} shows the continuous Sm-A -- Sm-C boundary as a straight line in $|\psi|^2$-$M$ space. At this point we make a notational distinction. In referring to the Sm-A -- Sm-C transition temperature generally (i.e. without distinguishing between continuous or first order) we use $T_{_{AC}}$. When referring specifically to either a continuous or a first order transition we use $T_0$ and $T_{1st}$, respectively. It is useful to work with a reduced temperature $t\equiv\frac{T}{T_0}-1$ which, near the continuous transition, can be related to $\tau$ via $\tau =1-\frac{|\psi_0(T)|^2}{|\psi_0(T_0)|^2} \approx p t$. Here we have Taylor expanded $|\psi_0(T)|$ near $T=T_0$. The dimensionless parameter $p= \left. -\frac{T_0}{|\psi_0(T_0)|^2} \frac{d |\psi_0(T)|^2}{d T} \right|_{T=T_0}>0$ can be thought of as a dimensionless measure of how rapidly the layering order changes with temperature. 

\subsubsection{Sm-A -- Sm-C tricritical point}
\label{Sm-A -- Sm-C Tricritical Point}

The dimensionless tricritical proximity parameter $\mu$ incorporates the renormalization of the $c^4$ term due to the coupling between biaxiality $\alpha$ and tilt $c$ (in the absence of such a coupling $\mu=1$) and depends on the amount of orientational and layering order. It is given by
\begin{eqnarray}
\mu(T)= 1-\frac{g}{2hq^2}\left(\frac{wM(T)}{gq^2|\psi(T)^2|}-1\right)^{-1}\;,
\label{mu}
\end{eqnarray}
where the temperature dependence of $\mu$ is a consequence of the temperature dependence of both $\psi$ and $M$. For de Vries materials, in which the orientational order $M$ varies very little with temperature in the Sm-A phase, the temperature dependence of $\mu$ in the Sm-A phase is due primarily to the temperature variation of the layering order $|\psi|$. Figure \ref{detailed phase diagram} shows the locus of $\mu=0$ in $|\psi|^2$-$M$ space. The nature of the transition is determined by the sign of $\mu_{_{AC}}\equiv\mu(T_{_{AC}})$, the value of $\mu$ at the zero-field Sm-A -- Sm-C transition. For $\mu_{_{AC}}>0$ (for small and large values of $|\psi|$ and $M$, respectively) the transition is continuous while for $\mu_{_{AC}}<0$ (for large and small values of $|\psi|$ and $M$, respectively) the transition is first order. When $\mu_{_{AC}}=0$ the quartic term vanishes and the transition is tricritical. As shown in Fig.~\ref{detailed phase diagram} the associated tricritical point ($|\psi_{_{TC}}|^2$, $M_{_{TC}})$ is located where the continuous Sm-A -- Sm-C boundary meets the locus of $\mu=0$. For de Vries materials with a virtually athermal $M$ the sign of $\mu_{_{AC}}$ is determined by the size of the system's orientational order. For a transition close to tricriticality, $\mu_{_{AC}}$ is most conveniently expressed as 
\begin{eqnarray}
\mu_{_{AC}} \approx m \left(\frac{M-M_{_{TC}}}{M_{_{TC}}}\right)\;,
\label{mu_AC}
\end{eqnarray}
where $m=1+ \frac{2 h q_0^2}{g}$ is a dimensionless constant. To lowest order in the coupling parameters, $M_{_{TC}} = \frac{m a_0 g^2}{2 h b w}$. The corresponding value of layering order at the tricritical point is $|\psi_{_{TC}}|=\sqrt{a/b}$. In previous models \cite{Huang&Viner} of the Sm-A -- Sm-C transition the parameter analogous to $\mu$ has been assumed to be independent of temperature. In our model, as discussed above, $\mu$ will vary with temperature via the temperature dependence of $|\psi(T)|$. For the time being we will use a constant $\mu$ approximation, $\mu(T)\approx\mu_{_{AC}}$, valid near the Sm-A -- Sm-C transition. In Section \ref{mu temperature dependence} we discuss in further detail the temperature dependence of $\mu$ and some of the related consequences for the electroclinic response.
\begin{figure}
\includegraphics[scale=1.7]{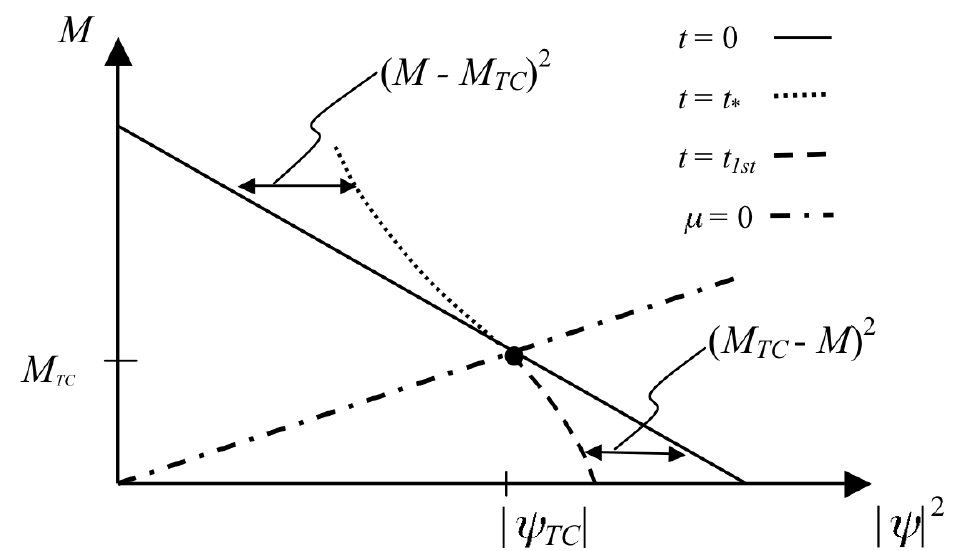}
\caption{The $t=0$ (solid), $\mu=0$ (dashed-dotted), $t=t_*\propto  -\left(M-M_{_{TC}}\right)^2$ (dotted) and $t= t_{_{1st}} \propto  \left(M_{_{TC}}-M\right)^2$ (dashed) locii in $|\psi|^2$-$M$ space. The corresponding phase diagram is shown in Fig.~\ref{phase diagram}(a). The continuous transition occurs for $\mu\geq0$ and at $t=0$. Thus, the tricritical point ($|\psi_{_{TC}}|^2$, $M_{_{TC}}$) is located at the intersection of the $t=0$ and $\mu=0$ locii. The first order Sm-A$^*$ -- Sm-C$^*$ transition occurs for $\mu<0$ and at $t>0$. The horizontal separation between the first order boundary and the extrapolated continuous boundary scales like $\left(M_{_{TC}}-M\right)^2$. Similarly, the separation between the continuous boundary and the tricritical crossover region at $t_*$ scales like $\left(M-M_{_{TC}}\right)^2$. The negative slope of the Sm-A$^*$ -- Sm-C$^*$ phase boundary implies that the smaller the value of $M$, the larger the value of $|\psi|$ at which the Sm-A$^*$ -- Sm-C$^*$ transition occurs. This is consistent with the observation that de Vries smectics generally have such unusually weak orientational order that their stabilization requires strong layering order, perhaps via micro-segregation.}
\label{detailed phase diagram}
\end{figure}

A commonly used \cite{Huang&Viner} measure of how close the continuous Sm-A -- Sm-C (or Sm-A$^*$ -- Sm-C$^*$) transition is to tricriticality is the magnitude of the reduced temperature, $|t_*|$, when $|r_c|=u_c^2/v_c$. Using this condition, it is straightforward to show that 
\begin{eqnarray}
t_*=-\kappa_1\left(\frac{M-M_{_{TC}}}{M_{_{TC}}}\right)^2 \;,
\label{tau_*}
\end{eqnarray}
where the dimensionless constant $\kappa_1=\frac{4h^2m^2}{3p a s}$. In the Sm-C phase, well within the corresponding temperature window $T_*<<T<T_0$, where $T_*=T_0(1-|t_*|)$, the quartic term $\propto c^4$ is important, and the behavior is $XY$-like. Sufficiently far outside this window, i.e. $T<<T_*$, it can be neglected, and the behavior of the system is tricritical. Figure \ref{detailed phase diagram} shows the corresponding crossover region in $|\psi|^2$-$M$ space, in which the system's behavior goes from being $XY$ to tricritical. The reduced temperature $t_*$ can be obtained \cite{Huang&Viner} from measurements of the specific heat at the continuous Sm-A -- Sm-C (or Sm-A$^*$ -- Sm-C$^*$) transition. Work has been done to relate the size of the reduced temperature window $|t_*|$ to system parameters, e.g. the width of the Sm-A phase \cite{Huang&Lien}. However, to the best of our knowledge, no one has yet investigated a possible relationship between the size of the reduced temperature window and the size of the orientational order $M$. The above expression for $t_*$ provides a prediction for such a relationship \cite{reduced temperature and order parameters}.

\subsubsection{First order Sm-A -- Sm-C transition}
\label{First Order Sm-A -- Sm-C Transition}

It was shown \cite{SaundersTCP}, that when the tricritical proximity parameter $\mu_{_{AC}}<0$, i.e. $M<M_{_{TC}}$, a first order Sm-A - Sm-C transition occurs at a value of $t$ given by $t_{_{1st}}=\frac{3}{16}|t_*|> 0$. As discussed in Ref.~\cite{SaundersTCP} the size of the latent heat at the first order Sm-A -- Sm-C transition is proportional to $\mu_{_{AC}}$ and thus, calorimetric studies can measure the proximity of the first order transition to the tricritical point. It is important to keep in mind that the first order Sm-A -- Sm-C will occur at $t>0$ and thus $T_{1st} >T_0$. Correspondingly, the value of layering order $|\psi|$ at the first order Sm-A -Sm-C boundary is smaller than would be necessary for a continuous Sm-A -- Sm-C transition. Fig.~\ref{detailed phase diagram}, shows an extrapolation of the continuous Sm-A -- Sm-C boundary in $|\psi|^2$-$M$ space for $M<M_{_{TC}}$. The difference between the layering at the extrapolated boundary and the first order Sm-A -- Sm-C boundary is proportional to $\left(M-M_{_{TC}}\right)^2$.

\subsection{The roles of orientational order and layering order in de Vries behavior and the nature of the Sm-A -- Sm-C transition}
\label{The roles of orientational order and layering order in de Vries behavior and the nature of the Sm-A -- Sm-C transition}

As shown in Ref.~\cite{SaundersTCP} de Vries behavior, i.e. an unusually small change in the layer spacing at the Sm-A -- Sm-C transition can be explained by unusually small orientational order and coupling parameters. The de Vries behavior, i.e. unusually rapid change, of the birefringence at the Sm-A -- Sm-C transition can be explained by proximity of the transition to a tricritical point. It has been experimentally observed \cite{HuangDV, Hayashi1, Collings} that several materials exhibiting de Vries behavior also have a Sm-A -- Sm-C transition that is close to a tricritical point. Our model implies that de Vries behavior and proximity of the Sm-A -- Sm-C transition to tricriticality can be connected by unusually small orientational order. Indeed, it has been observed that de Vries materials  do have unusually small orientational order. Consequently it has been argued \cite{de Vries review, S Lagerwall} that stabilization of materials with such small orientational order must be provided by unusually strong layering order, perhaps via micro-segregation. The phase diagram in $|\psi|^2$-$M$ space, shown in Fig.~\ref{phase diagram}, is consistent with such an argument; the negative slopes of both the continuous and first order phase boundaries mean that systems with smaller orientational order require larger layering order to make the transition from the Sm-A phase to the Sm-C phase. 

To the best of our knowledge, no direct measurement of the layering order in de Vries materials has been published. We believe such measurements would be valuable in understanding the role that layering order plays in driving the Sm-A -- Sm-C (or Sm-A$^*$ -- Sm-C$^*$ ) transition, as well as the nature of the transition (i.e. continuous, tricritical, or first order) and how de Vries-like the system is. While direct measurements of the layering have not been reported, there is published data \cite{Krueger} on the width of the Sm-A$^*$ phase in a homologous series of hexyl lactates ($\it n$HL) exhibiting Sm-A$^*$ -- Sm-C$^*$  transitions that range from conventional to de Vries-like. It is found that the temperature width of the Sm-A$^*$ phase window increases as the system becomes more de Vries-like. Making the conventional assumption that the layering order at the Sm-A$^*$ -- Sm-C$^*$  transition is a monotonically increasing function of the temperature width of the Sm-A$^*$ phase, this data is consistent with our model. However, one must be careful in making this assumption for systems (e.g. de Vries materials) that have first order isotropic (Iso) -- Sm-A (or Sm-A$^*$) transitions where the layering does not necessarily grow continuously from zero. For example, it could be possible that the layering order at the Iso -- Sm-A (or Sm-A$^*$) transition is larger in systems with smaller orientational order. Thus, the layering at the Iso -- Sm A (or Sm-A$^*$) transition may already be large enough so that it is not necessary to have a wider temperature window for the Sm-A (or Sm-A$^*$) phase \cite{layering T window assumption}. This is another reason that a systematic experimental investigation of the layering and orientational order in these systems would be valuable.

\section{Incorporating the effects of chirality and external fields to the free energy density}
\label{Incorporating the effects of chirality and external fields to the free energy density}

Having set up the nonchiral, zero-field free energy we next add terms to reflect the presence of chirality and an externally applied field. The most important such term is the one which models the electroclinic interaction of the molecules with the applied electric field ${\bf E}$. To lowest order in the orientational and layering order parameters it is
\begin{eqnarray}
f_{_{EC}}= e\epsilon_{ijk} q_i q_l |\psi|^ 2 E_j Q_{lk}\approx e q^2 M |\psi|^ 2 {\bf \hat z} \cdot ({\bf E} \times {\bf c} ) \;,
\label{f_EC}
\end{eqnarray}
where $\epsilon_{ijk}$ is the Levi-Cevita symbol and the Einstein summation convention is implied. In coupling the electric field directly to the tilt ${\bf c}$, instead of via the electrostatic polarization $\bf P$, we are making the standard assumption that ${\bf P} \propto {\bf \hat z}\times{\bf c}$ \cite{eclinic coupling}.  The coefficient $e$ depends on the strength of the electrostatic coupling between the field and the molecules. This in turn depends on amount of chirality in the system and for a racemic mixture $e=0$. Here we take $e>0$; switching the handedness, e.g. left to right, of the molecules simply switches the sign of $e$. In making the approximation in Eq.~(\ref{f_EC}) above, we include only the lowest order contribution  of tilt $c$ from the orientational order tensor ${\cal Q}$ and we neglect the biaxial part of ${\cal Q}$. It can be shown that close to the tricritical point the coupling between the field and biaxiality is negligible. 

In order to make the model manageable, we also omit other contributions, each of which lead only to secondary, less important effects. The first of these is the nonchiral coupling of the system to the electric field which would contribute terms such as $E_iE_jQ_{ij}$. All such terms scale like $E^2$ and in the limit of small field can be shown to be much smaller than the electroclinic term in Eq.~(\ref{f_EC}) above, which scales linearly with $E$.  

We also assume a spatially uniform tilt ${\bf c}$ and thus ignore a second group of contributions involving spatial variations in ${\bf c}$, including manifestly chiral terms that depend on the sign of ${\bf \nabla} \times {\bf c}$. We have analyzed the difference that such terms make to our model. One zero-field effect of these terms is to shift the location of the Sm-A$^*$ -- Sm-C$^*$  phase boundary, by renormalizing the coefficients in the free energy expression, Eq.~(\ref{f_c}), for $f_c$. In particular, increasing the chirality of the system lowers the quadratic coefficient, $r_c$, the effect of which is to increase the Sm-A$^*$ -- Sm-C$^*$  transition temperature. Increasing the chirality also lowers the value of the quartic coefficient $u_c$, thus driving a continuous transition towards tricriticality or a first order transition away from tricriticality. The behavior of the layer spacing and birefringence are also somewhat affected via the renormalization of these coefficients. However, in the limit (which we assume throughout) of small orientational order and small couplings between layering and orientational order parameters, the renormalization of these coefficients is negligible. Thus, the zero-field behavior of the chiral system should essentially be the same as described for the nonchiral system.

The absence of terms involving spatial variations in ${\bf c}$ also precludes the possibility of a superstructure involving a spatial modulation of ${\bf c}$, which in the zero-field Sm-C$^*$ phase would be helical. In the past \cite{Bahr Heppke} the assumption of a spatially uniform tilt has been justified by consideration of electric field strength above that necessary to unwind a helical superstructure. However, it is not obvious that a helical superstructure would form when the tilt is electrically induced (as opposed to spontaneously developing at the zero-field Sm-A$^*$ -- Sm-C$^*$  transition.) For example, it has been shown \cite{2d Eclinic effect} that in a two-dimensional Sm-A$^*$ film, the electroclinic effect can lead to a spatially uniform tilt at small and large fields and to a modulated tilt for fields of intermediate strength. To the best of our knowledge the situation for three-dimensional Sm-A$^*$ systems has yet to be analyzed, although we plan to do so in the near future. It should be pointed out that one proposed explanation \cite{Meyer Pelcovits} for the strong electroclinic effect in de Vries materials is that the Sm-A$^*$ phase is actually a Sm-C$^*$ phase that is made up of an ordered array of disclination lines and walls, and thus assumes a strong spatial modulation of the tilt in the Sm-A$^*$ phase. We do not explore that possibility here. 

In summary, because we are interested primarily in the electroclinic effect and do not wish to overburden the model with less important secondary effects,  the only extra term we add to our nonchiral model is that given in Eq.~(\ref{f_EC}). 

\section{Response of Tilt}
\label{Response of Tilt}

In this section we explore the response of the tilt order parameter ${\bf c}$ to an externally applied electric field ${\bf E}$. Of particular interest is the response near the tricritical point shown in Fig.~\ref{phase diagram}. As shown in Fig.~\ref{Eigenvectors} we take the field to point in the ${\bf \hat y}$ direction, so that the free energy is minimized by a tilt in the ${\bf \hat x}$ direction, i.e. ${\bf c} = c {\bf \hat x}$ and $f_{_{EC}}=-b q^2 M |\psi|^ 2Ec$. The magnitude $c$ of the tilt induced by the applied field can be determined using the tilt portion of the free energy, $f_{c} +f_{_{EC}}$. Minimizing this free energy with respect to the tilt $c$ one obtains the following relationship between $c$ and $E$:
\begin{eqnarray}
E= \alpha_e c + \beta_e c^3 + \gamma_e c^5\;,
\label{E(c)}
\end{eqnarray}
where the electroclinic coefficients $\alpha_e$, $\beta_e$ and $\gamma_e$ are given by
\begin{eqnarray}
\alpha_e &=&\frac{3 a p t}{e}\;,
\label{alpha}
\end{eqnarray}
\begin{eqnarray}
\beta_e &=& \frac{9\mu h q^2 M}{e} \; ,
\label{beta}
\end{eqnarray}
\begin{eqnarray}
\gamma_e&=& \frac{81s q^4 M^2}{4e} \;,
\label{gamma}
\end{eqnarray}
where the reader is reminded that the tricritical paramter $\mu$ generally depends on temperature via its dependence on orientational ($M$) and layering order ($|\psi|$), given in Eq.~(\ref{mu}). As discussed in Section \ref{Sm-A -- Sm-C Tricritical Point} the orientational order of the Sm-A$^*$ phase in de Vries materials varies very little with temperature, so the temperature dependence of $\mu$ in the Sm-A$^*$ phase is due primarily to the temperature variation of the layering order $|\psi|$.  The relationship Eq.~(\ref{E(c)}) between $E$ and $c$ is analogous \cite{tilt polarization footnote} to that derived by Bahr and Heppke in their analysis of a field-induced critical point near the Sm-A$^*$ -- Sm-C$^*$ transition \cite{Bahr Heppke}. There are, however, a couple of distinctions that should be pointed out. The first is motivation. In Ref.~\cite{Bahr Heppke} the primary motivation was to establish the existence of and to analyze a line of first order Sm-A$^*$ -- Sm-C$^*$ transitions in the temperature-field plane that terminates at a critical point. Our motivation is to model and explain the unusually large electroclinic response of de Vries materials. It will be shown that this can be done by analyzing Eq.~(\ref{E(c)}) in a similar manner to Ref.~\cite{Bahr Heppke}.  

A second, related, distinction is that, as a result of starting with a generalized Landau theory in terms of orientational and layering order parameters, we can relate our coefficients $\alpha_e$, $\beta_e$, $\gamma_e$ to the strengths of orientational order (and hence birefringence) and layering order, as well as the layer spacing (via $q$) in the system. Of particular interest is the origin of a negative quartic coefficient ($\beta_e<0$ in Eq.~(\ref{E(c)}) above) which is necessary for a field-induced first order Sm-A$^*$ -- Sm-C$^*$ transition. In Ref.~\cite{Bahr Heppke}, this was assumed (justifiably) on the basis of the existence of a zero-field first order Sm-A$^*$ -- Sm-C$^*$ transition. Here, a negative quartic coefficient can be explained as resulting from sufficiently weak orientational order, which, as discussed in Section \ref{The roles of orientational order and layering order in de Vries behavior and the nature of the Sm-A -- Sm-C transition} necessitates strong layering order, in order to stabilize the system. Thus, our generalized Landau theory shows that an unusually strong electrical response of the tilt can be explained as resulting from a combination of weak orientational order and strong layering order, which makes the quartic coefficient $\beta_e$ either positive and small (corresponding to a continuous zero-field Sm-A$^*$ -- Sm-C$^*$ transition that is near a tricritical point) or negative (corresponding to a first order zero-field Sm-A$^*$ -- Sm-C$^*$ transition). A related distinction between this analysis and that of Bahr and Heppke, is that our quartic coefficient, $\beta_e$ depends on temperature via the temperature dependence of $\mu$. We will next analyze the electroclinic response implied by Eq.~(\ref{E(c)}). 

\subsection{Electroclinic response near the continuous zero-field Sm-A$^*$ -- Sm-C$^*$ transition}

We begin our analysis by approximating the tricritical proximity parameter $\mu$ as being temperature independent, i.e. $\mu(T)\approx\mu_{_{AC}}$, which is valid sufficiently close to the Sm-A$^*$ -- Sm-C$^*$ transition. The effect of $\mu$'s temperature dependence will be discussed in Section \ref{mu temperature dependence}. For $\mu_{_{AC}}>0$,  corresponding to a continuous zero-field Sm-A$^*$ -- Sm-C$^*$ transition, $\beta_e>0$. For such systems, the response of the tilt $c$ to an applied field $E$ is continuous. Additionally, as shown in Fig. {\ref{c vs E, beta>0}, the susceptibility $\chi=\frac{\partial c}{\partial E}$, gets smaller with increasing field. Its largest value, at $E=0$, is $\chi_0(T)=\alpha_e(T)^{-1}$, which diverges as the system approaches the continuous zero-field Sm-A$^*$ -- Sm-C$^*$ transition at $\alpha_e(T_0) =0$, a standard result for continuous transitions.
\begin{figure}
\includegraphics[scale=0.8]{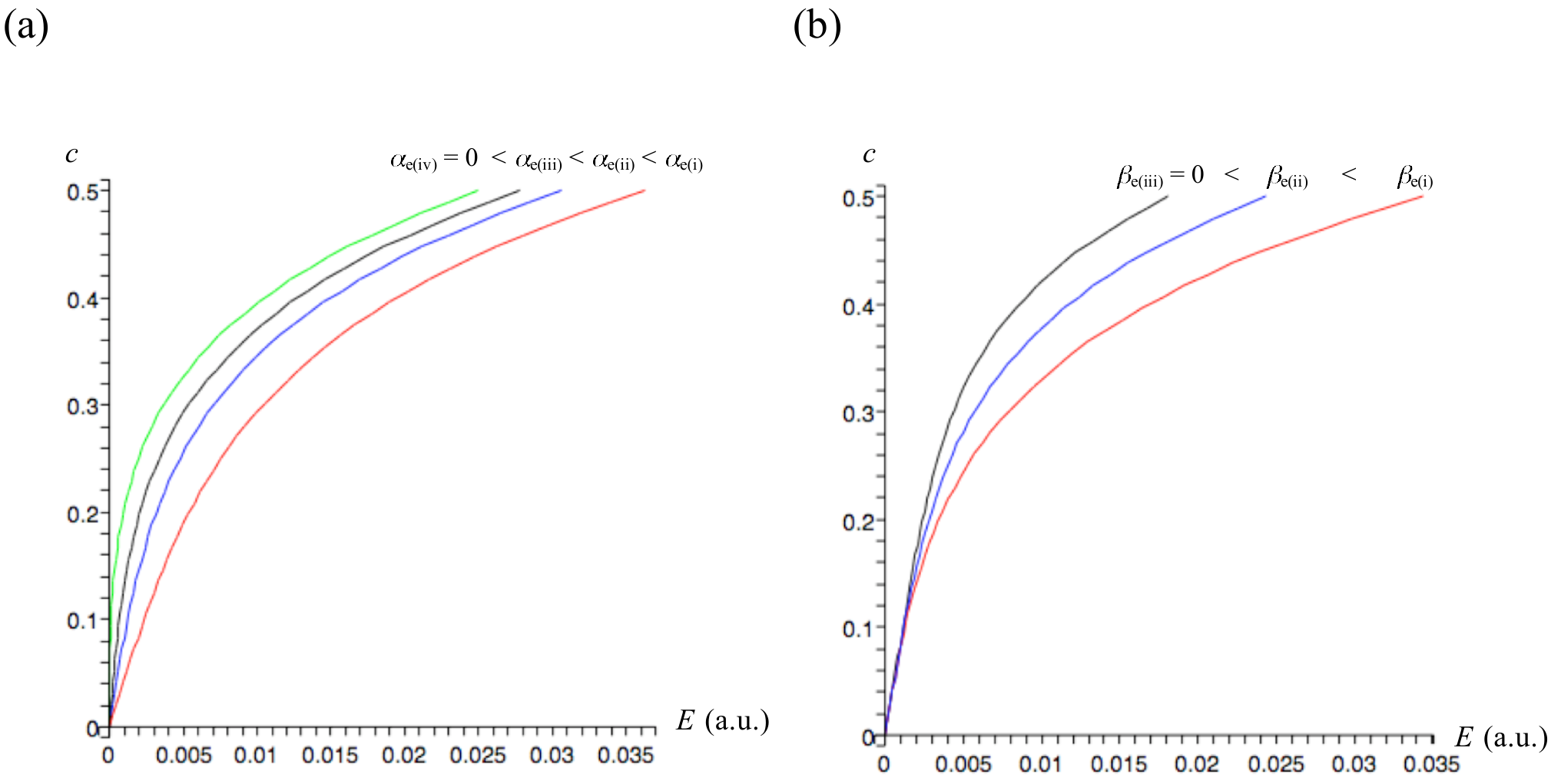}
\caption{Response curves $c(E)$ for systems with $\beta_e\geq0$, i.e. systems with continuous Sm-A$^*$ -- Sm-C$^*$ transitions. The curves show the electrically induced tilt $c$ due to the application, in the Sm-A$^*$ phase, of a field of magnitude $E$. The curves were obtained using Eq.~(\ref{E(c)}). Since we are primarily interested in the evolution of the shape of the curves we do not specify units for $E$ (i.e., we use arbitrary units, a.u.). 
(a) A set of curves for fixed $\gamma_e=0.4$, $\beta_e=0.1$ (and thus, fixed $\mu_{_{AC}}>0$) and different values of $\alpha_e \propto  t \geq 0$. The different values of $\alpha_e$ correspond to different values of reduced temperature values $t \geq 0$, and thus to different values of $T\geq T_{_{AC}}$. (i) $\alpha_e=0.0225$, (ii) $\alpha_e= 0.01125$, (iii) $\alpha_e=0.005625$, (iv) $\alpha_e=0$. The susceptibility $\chi = \frac{dc}{dE}$ is largest at $E=0$, and monotonically decreases as $E$ is increased. The response increases as temperature, $T$, is lowered towards the Sm-A$^*$ -- Sm-C$^*$ transition temperature, $T_{_{AC}}$, with the zero-field susceptibility $\chi_{_0}$ diverging as $T$ approaches $T_{_{AC}}$ (or equivalently, as $\alpha_e$ approaches zero).  
(b) A set of curves for fixed $\alpha_e=0.01125$ (and thus fixed reduced temperature $t>0$), $\gamma_e=0.4$ and different values of $\beta_e \propto  \mu_{_{AC}}\geq0$. The different values of $\beta_e\geq0$ imply varying degrees of proximity of the continuous Sm-A$^*$ -- Sm-C$^*$ transition to a tricritical point. (i) $\beta_e= 0.13$, (ii) $\beta_e= 0.05$, (iii) $\beta_e=0$. The response is larger for systems with smaller $\beta_e$  (and thus, smaller $\mu_{_{AC}}$).}
\label{c vs E, beta>0}
\end{figure}
The response at the  Sm-A$^*$ -- Sm-C$^*$ transition for small fields is $c\propto E^{\frac{1}{\delta}}$, with $\delta=3$ away from tricriticality and $\delta=5$ at the tricritical point. 

It is interesting to consider how the response ($c$) at fixed reduced temperature ($t>0$) and field ($E>0$) is affected by  lowering $\mu_{_{AC}}$ towards zero, i.e. driving the continuous transition to tricriticality. It is straightforward to show that
\begin{eqnarray}
\frac{\partial c}{\partial\mu_{_{AC}}}&=& - \frac{9hq^2Mc^3\chi}{e}<0\;,
\label{susceptibilty}
\end{eqnarray}
which, as expected, shows that the response, at fixed $E$ and $t$ should be larger for systems with smaller $\mu_{_{AC}}$, i.e. systems in which the orientational order is small ($M\gtrsim M_{_{TC}}$). This is shown graphically in Fig.~\ref{c vs E, beta>0}(b) and is reminiscent of an experimentally obtained comparison \cite{de Vries review} of electroclinic responses for a homologous series of hexyl lactates ($\it n$HL), with each response being measured at the same reduced temperature. The response is observed to be larger for small $n$ vaues. The compounds have zero-field continuous Sm-A$^*$ -- Sm-C$^*$  transitions that range from conventional to de Vries-like \cite{Krueger}.  We speculate that if one were to measure the proximity of each compound's transition to a tricritical point, one would find that 9HL's and 12HL's transitions are closest and furthest respectively, i.e. $0<\mu_{_{{AC}_{_{9HL}}}}<\mu_{_{{AC}_{_{12HL}}}}$.

\subsection{Electroclinic response near the first order zero-field Sm-A$^*$ -- Sm-C$^*$ transition}

Next we consider the response when the tricritical proximity parameter $\mu_{_{AC}}<0$ (and thus $\beta_e<0$) corresponding to a first order zero-field Sm-A$^*$ -- Sm-C$^*$ transition. As shown in Fig.~\ref{c vs E, beta<0}, for large reduced temperatures $t$ the response is continuous and will show a positive curvature ($\frac{d^2 c}{d E^2}>0$) at small fields followed by a negative curvature ($\frac{d^2 c}{d E^2}<0$) at large fields. The positive curvature has been referred to in the literature (e.g. in Ref.~\cite{Clark}) as ``superlinear growth". For sufficiently small temperatures the response curve is $\cal S$ shaped (i.e. has a portion with negative slope) and there is a jump in the tilt as the electric field is increased from zero. Thus, an unusually strong, discontinuous, electroclinic effect will be exhibited by systems with sufficiently small orientational order $M< M_{_{TC}}$.
\begin{figure}
\includegraphics[scale=0.8]{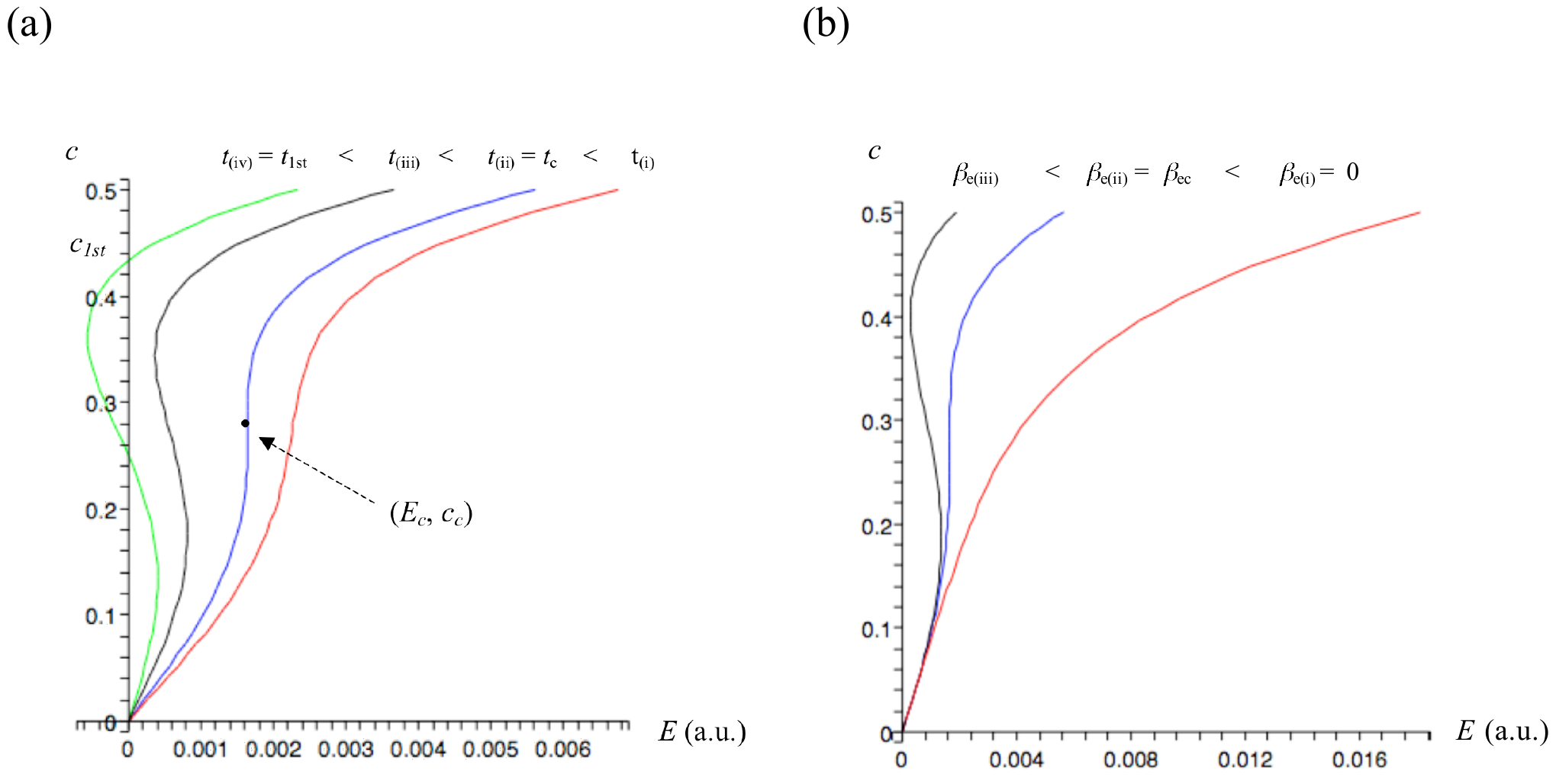}
\caption{Response curves $c(E)$ for systems with $\beta_e<0$, i.e. systems with first order Sm-A$^*$ -- Sm-C$^*$ transitions. The curves show the electrically induced tilt $c$ due to the application, in the Sm-A$^*$ phase, of a field of magnitude $E$. The curves were obtained using Eq.~(\ref{E(c)}). Since we are primarily interested in the evolution of the shape of the curves we do not specify units for $E$ (i.e., we use arbitrary units, a.u.). 
(a) A set of curves for fixed $\gamma_e=0.4$, $\beta_e=-0.1$ (and thus, fixed $\mu_{_{AC}}<0$) and different values of $\alpha_e$. The values of $\alpha_e$ are most usefully expressed in terms of  $\alpha_{ec}\equiv \frac{9}{20}\frac{\beta_e^2}{\gamma_e}=0.00125$, which is the value of $\alpha_e$ at the critical reduced temperature $t_c$. Below this value the curves become ${\cal S}$ shaped. Since $\alpha_e\propto t$, the ratio $t/t_c$ is the same as $\alpha_e/ \alpha_{ec}$ and we label the curves according to the value of $t$ in terms of $t_c$ : (i) $t=1.2 t_c$, (ii) $t= t_c$, (iii) $t=0.65 t_c$, (iv) $t=t_{1st}=\frac{5}{12}t_c$. For $t>t_c$ the response is continuous but ``superlinear", with $\frac{d^2 c}{d E^2}>0$ at small $E$ followed by $\frac{d^2 c}{d E^2}<0$ at large $E$. At $t=t_c$ the response has divergent susceptibility (i.e., slope) and curvature at $E_c$ and $c_c$, indicated with a dot. The curves are now ${\cal S}$ shaped which implies a discontinuous response. 
(b) A set of curves for fixed $\alpha_e=0.01125$ (and thus, fixed reduced temperature $t>0$), $\gamma_e=0.4$ and different values of $\beta_e \propto  \mu_{_{AC}}\leq0$. Different values of $\beta_e$ imply varying degrees of proximity of the first order Sm-A$^*$ -- Sm-C$^*$ transition to a tricritical point. The values of $\beta_e$ are given in terms of $\beta_{ec} \equiv -\sqrt{\frac{20\alpha_e\gamma_e}{9}}$, the value of $\beta_e$ below which the curves become ${\cal S}$ shaped. (i) $\beta_e=0$, (ii) $\beta_e= \beta_{ec}$, (iii) $\beta_e=1.3\beta_{ec}$. Making $\beta_e$ (and thus, $\mu_{_{AC}}$) more negative would increase the temperature window $T_c-T_{1st}$ so that the discontinuous response occurs further away from the Sm-A$^*$ -- Sm-C$^*$ transition.}
\label{c vs E, beta<0}
\end{figure}

We define $t_c$ as the value of reduced temperature below which the response curve, $c(E)$, exhibits a negative slope, and hence a discontinuity in the response. As shown in Fig.~\ref{c vs E, beta<0}, at $t=t_c$, the curve has divergent slope and curvature at $E_c$ and $c_c$. Thus, the values $t_c$, $E_c$ and $c_c$ are specified by $\frac{d E}{d c}|_{_{t_c,c_c}}=\frac{d^2 E}{d c^2}|_{_{t_c,c_c}}=0$ and $E_c=E(t_c,c_c)$. At the level of the mean-field theory presented here and elsewhere \cite{Bahr Heppke} the associated critical point $(t_c,E_c,c_c)$ is analogous to the liquid-vapor critical point. However, it has been pointed out \cite{Prost} that when fluctuations are included the universality class of this critical point is distinct from that of the liquid-vapor critical point.
It is straightforward \cite{Bahr Heppke} to calculate the critical values $(t_c,E_c,c_c)$. We rederive these values, primarily with a view to presenting them in terms of the degree of orientational order in the system. We also provide extra details that might be useful in experimentally investigating whether the strong response of de Vries materials is indeed a result the proximity of the Sm-A$^*$ -- Sm-C$^*$ tricritical point. The value of the critical reduced temperature is found to be $t_c=\frac{12}{5}t_{1st}$, where $t_{1st}$ is defined in Section \ref{First Order Sm-A -- Sm-C Transition}. Keeping in mind the fact that the first order transition occurs at $t_{1st}>0$, one can find the temperature difference between $T_{1st}$ and $T_c$:
\begin{eqnarray}
\frac{T_c-T_{1st}}{T_{1st}}&\approx& \frac{21}{80}\kappa_1\left(\frac{M_{_{TC}}-M}{M_{_{TC}}}\right)^2\;,
\label{T_C}
\end{eqnarray}
where the dimensionless constant $\kappa_1=\frac{4h^2m^2}{3p a s}$ was defined earlier in Section\ref{Sm-A -- Sm-C Tricritical Point} and the approximation applies close to tricriticality where $M\lesssim M_{_{TC}}$. The value of $c_c$ is found to be
\begin{eqnarray}
c_c=\sqrt{\frac{2}{5}}c_{1st}\propto\left(\frac{M_{_{TC}}-M}{M_{_{TC}}}\right)^{\frac{1}{2}}\;,
\label{c_c ito c_{1st}}
\end{eqnarray}
where $c_{1st}$ is the size of the jump in the tilt order parameter at the zero-field transition, which is found \cite{SaundersTCP}  to be $c_{1st}=\sqrt{\frac{3|\beta_e|}{4\gamma_e}}$. The above equation implies that the ratio $\frac{c_c}{c_{1st}}=\sqrt{\frac{2}{5}}$ should hold for any system, a prediction that should be straightforward to test experimentally. Lastly, we find 
\begin{eqnarray}
E_c=\frac{8}{15}\alpha_e(t_c)c_c\propto\left(\frac{M_{_{TC}}-M}{M_{_{TC}}}\right)^{\frac{5}{2}}\;,
\label{E_C}
\end{eqnarray}
where $\alpha_e(t)$ is given by Eq.~(\ref{alpha}). Using the fact that $\alpha_e(T)=\chi_0(T)^{-1}$, we define the following combination 
\begin{eqnarray}
\lambda_c = \frac{\chi_0(T_c) E_c}{c_c}\;,
\label{lambda_c}
\end{eqnarray}
where $\chi_0(T_c)$ is the value of the zero-field susceptibility at $T=T_c$. Together, Eqs. (\ref{E_C}) and (\ref{lambda_c}) predict that $\lambda_c=8/15$ for {\it every} material that has a first order Sm-A$^*$ -- Sm-C$^*$ transition. It would be interesting to investigate experimentally whether this is accurate for de Vries materials with first order Sm-A$^*$ -- Sm-C$^*$ transitions. If so, it would indicate that the mean-field theory described here is suitable to describe the strong electro-optic response of de Vries type materials.

\subsection{The effects of the temperature dependence of the tricritical proximity parameter $\mu$ on the electroclinic response near the Sm-A$^*$ -- Sm-C$^*$ transition}
\label{mu temperature dependence}

In previous models \cite{Huang&Viner} of the Sm-A -- Sm-C (and Sm-A$^*$ -- Sm-C$^*$) transitions the parameter analogous to $\mu$ has been assumed to be independent of temperature. In our model $\mu(T)$, given by Eq.~(\ref{mu}), will vary with temperature via the temperature dependence of $|\psi(T)|$ and to a lesser degree $M(T)$. From Eq.~(\ref{mu}) it can be seen that $\mu(T)$ decreases if $M(T)$ and $|\psi(T)|$ decrease and increase, respectively. We have argued here and elsewhere \cite{Saunders1, SaundersTCP}, that in de Vries materials the system is driven towards the Sm-C$^*$ phase as the layering ($|\psi|$) increases with decreasing temperature. Additionally, the nonmonotonicity of $M(T)$, that is both predicted by our model and observed experimentally \cite{Lagerwall, Manna} in de Vries materials, implies that $M(T)$ decreases as the Sm-A$^*$ -- Sm-C$^*$ transition is approached from above. Thus, each of these effects causes $\mu(T)$ to decrease towards $\mu_{_{AC}}$ as the Sm-A$^*$ -- Sm-C$^*$ transition is approached from above. 

As discussed in the preceding two sections, decreasing $\mu(T)$ leads to a strengthening of the electrical response of the tilt. Thus, we speculate that the electroclinic response in de Vries materials is further strengthened by the thermal behavior of the layering and orientational order. It would be interesting to extract the temperature dependence of $\mu(T)$ (perhaps through fitting the response curves at different temperatures) to see if it does have a temperature dependence and, if so, whether it decreases as the Sm-A$^*$ -- Sm-C$^*$ transition is approached from above.  

There may also be an observable feature associated with the temperature dependence of $\mu(T)$ and the nonmonotonicity of $M(T)$. It has been predicted \cite{Saunders1} and observed \cite{Manna} that $M(T)$ can have a maximum within the Sm-A$^*$ phase. This would correspond to a birefringence that increases  with decreasing temperature (after the system has entered the Sm-A$^*$ phase from the isotropic phase) before reaching a maximum at $T_{\max}$ and then decreases as the Sm-C$^*$ phase is approached. For systems in which this is the case, as $T$ is lowered through $T_{\max}$ the decrease in $\mu(T)$ would become more rapid once $M(T)$ begins to decrease. If this were so, there may be an associated anomaly in the electroclinic response as $T$ is lowered through $T_{\max}$.

\section{Response of the birefringence and layer spacing to an electric field applied to the Sm-A$^*$ phase}
\label{Response of Birefringence and Layer Spacing}

Having analyzed how the tilt order parameter, ${\bf c}$, will respond to an electric field ${\bf E}$ being applied to the Sm-A$^*$ phase, we now investigate how the birefringence, $\Delta n$, and layer spacing, $d$, are simultaneously affected. We do this with a view to providing insight into the response of birefringence and layer spacing for de Vries materials in particular. First we summarize the main experimental observations \cite{de Vries review}. In de Vries materials the response of the tilt is unusually strong, which as discussed above, can be explained by an unusually small orientational order which leads to a Sm-A$^*$ -- Sm-C$^*$ transition that is either continuous and close to tricriticality or first order. The response of the birefringence (which is proportional to the orientational order in the system) is also unusually strong. However, the contraction, i.e. fractional change $\Delta_d$, of the layer spacing $d$ associated with the tilt is unusually small. The combination of a large response in the tilt and birefringence and a small contraction of the layer spacing is technologically desirable. The unusually small contraction of the layers eliminates buckling of the layers and the associated chevron defects which lead to unwanted striping in ferroelectric liquid crystal displays.

Another noteworthy experimental observation is the scaling of the birefringence response with tilt response.  The tilt $c(E)$ and the birefringence $\Delta n(E)$ each scale nonlinearly with applied field $E$, and the shape of the nonlinear curves change significantly as temperature is varied. However, a parametric plot of  $\Delta n(E)$ vs $c^2(E)$, is very close to being linear. Remarkably, this linear scaling seems to hold regardless of the nature (i.e. continuous, tricritical, or first order) of the transition \cite{continuous response, discontinuous response}. Additionally, the slope of this linear scaling varies very little with temperature. There does not seem to be any published parametric plots of  $\Delta_d(E)$ as a function $c^2(E)$. As discussed in more detail below, we predict that while $\Delta_d(E)$ will scale nonlinearly with applied field it will scale linearly with $c^2(E)$. The slope of this linear scaling is proportional to the orientational order and will thus be unusually small in systems with unusually small orientational order. Unlike the birefringence we predict that the slope of the absolute change in layer spacing $\Delta d(E)$ (as opposed to fractional change $\Delta_d(E)$) vs $c^2(E)$ will not be weakly temperature dependent.

In what follows we first investigate the response of the birefrigence to an applied electric field. The general methods described here are also applied to investigating the response of the layer spacing. 

\subsection{Response of the birefringence to an electric field applied to the Sm-A$^*$ phase}
\label{Response of Birefringence}

In analyzing the response of the birefringence we use the fact that birefringence is proportional to the orientational order in the system and find the change in orientational order due to an applied field. We define the zero-field orientational order as $M_{_{E=0}}$. It is important to note that this differs from $M_0$, given after Eq.~(\ref{f_Psi}), which is defined as the zero-field orientational order in the absence of coupling between orientational order and layering. As discussed in the analysis of the nonchiral zero-field model \cite{SaundersTCP}, the effect of the coupling of the orientational order to layering order is to increase the orientational order above its zero coupling value $M_0$. Here, with our chiral model, we are focussing on the {\it additional} effect on orientational order, due to the application of an electric field. Thus, we use the notation $M_{_{E=0}}$ to represent the the zero-field orientational order, which includes the increase due to the zero-field coupling of orientational order to layering. This means that $M_{_{E=0}}>M_0$. As was shown in Ref.~\cite{Saunders1} $M_{_{E=0}}$ is a non-monotonic function of temperature. As temperature is lowered towards $T_{_{AC}}$, $M_{_{E=0}}$ {\it decreases} (albeit weakly), a feature which, while unusual, has nonetheless been observed experimentally \cite{Lagerwall, Manna}. Upon entry to the Sm-C phase, $M_{_{E=0}}$ increases with decreasing temperature. For continuous transitions the rate of increase is larger the closer the transition is to tricriticality and for first order transitions the increase is larger the further transition is from tricriticality. 

We define $\Delta_{M_E}$ as the fractional change in the orientational order, due to the application of an electric field, i.e., $M=M_{_{E=0}}(1+\Delta_{M_E})$. The response $\Delta_{M_E}$ is obtained by minimizing the free energy with respect to $\Delta_{M_E}$. This is made tractable by assuming that $\Delta_{M_E}$ is small and expanding the free energy to quadratic order in $\Delta_{M_E}$. Details of the analysis are given in the Appendix. We find that within the Sm-A$^*$ phase, for small $t$ and $\mu_{_{AC}}$, i.e., close to a Sm-A$^*$ -- Sm-C$^*$ transition which is close to tricriticality, the fractional change in orientational order is given by:
\begin{eqnarray}
\Delta_{M_E}=\frac{3m}{2\gamma_{_M}} g q_{_{E=0}}^2|\psi|^ 2\left[1- {\cal{O}}\left( \frac{c^2(E)}{c_{_M}^2}\right) \right] c^2(E)  \;,
\label{Delta M_E}
\end{eqnarray}
where $m=1+ \frac{2 h q_{_{E=0}}^2}{g}$ is a dimensionless constant and $\gamma_{_M}= \left. d^2 f_M /d M^2\right|_{M=M_0}$, where $f_M$ is given in Eq.~(\ref{H_M}). The zero-field layering wavevector,  $q_{_{E=0}}$ is distinct from the bare $q_0$, in that it includes the effects of the zero-field coupling between orientational and layering orders. The dimensionless parameter $c_{_M}$ can be thought of as the value of $c$ where the scaling of $\Delta_{M_E}$ with $c$ crosses over from being quadratic to quartic. We define $c_{_M}$ in the Appendix and show it to be $ {\cal{O}}(1)$, which makes the quartic contribution negligible in our theory, where it is assumed that $c\ll1$. It should also be pointed that the  largest experimentally measured values of $c$, obtained for large fields,  are on the order of $c_{max} \approx 0.5$ (corresponding to $c_{\max}=\sin(\theta_{\max})$, where $\theta_{\max}\approx30^0$).  Thus, at all but the largest values of $c$ the scaling of $\Delta_{M_E}$ with $c$ is quadratic, which is consistent with experiment. Most importantly, the above result, Eq.~(\ref{Delta M_E}), is valid for both continuous and discontinuous $c(E)$ response curves. Of course, the linear scaling of $\Delta_{M_E}$ with $c^2(E)$, implied by Eq.~(\ref{Delta M_E}), means that if there is a strong or discontinuous response of tilt $c$ to applied field $E$, there will be a correspondingly strong response of $M$, and hence birefringence, to applied field. This is also consistent with experiment.

Having shown that the change in orientational order (and hence birefringence) scales linearly with $c^2(E)$, we next consider the slope of this scaling, in particular its temperature dependence which, as discussed above, is experimentally observed to be weak. In most published work (e.g. Refs. \cite{Collings, Clark}) that has analyzed the change in birefringence as a function of tilt, it is the absolute change rather than the fractional change of briefringence that is considered. In our theory this corresponds to the absolute change in orientational order $\Delta M(E) = M-M_{_{E=0}}$ which is given by:
\begin{eqnarray}
\Delta M \propto M_{_{E=0}}(T) q^2_{_{E=0}}(T) |\psi(T)|^2 c^2(E)  \;,
\label{Absolute Delta M_E}
\end{eqnarray}
where we have used $\Delta M =M_{_{E=0}}\Delta_{M_E}$ and in going from Eq.~(\ref{Delta M_E}) to Eq.~(\ref{Absolute Delta M_E}) we have kept only the leading order temperature dependence (which we now display explicitly)  of the $c^2(E)$ prefactor. Thus, the temperature dependence of the slope of $\Delta M(E) $ vs $c^2(E)$ is determined by the temperature dependent combination $\sigma(T)=M_{_{E=0}}(T) q^2_{_{E=0}}(T) |\psi(T)|^2$. Since $M_{_{E=0}}(T)$,  $q_{_{E=0}}(T)$ and $|\psi(T)|$ each remain finite within the Sm-A$^*$ phase, both $\sigma(T)$ and the slope will also remain finite. In particular there will be no dramatic change in the slope as the Sm-A$^*$ -- Sm-C$^*$ transition is approached from above. Given that the temperature dependence of $M_{_{E=0}}(T)$ is weak, any change in the slope should be due to a change in the combination $q^2_{_{E=0}}(T) |\psi(T)|^2$. We have already argued that $|\psi(T)|$ increases monotonically as the Sm-A$^*$ -- Sm-C$^*$ transition is approached from above. It is generally observed experimentally that as the Sm-A$^*$ -- Sm-C$^*$ transition is approached from above, there is a monotonic dilation of the layer spacing, which corresponds to a monotonic decrease in $q_{_{E=0}}(T)$. Thus, we speculate that the temperature changes in $q^2_{_{E=0}}(T)$ and  $|\psi(T)|^2$ offset each other which leads to only a weak temperature dependence of the slope of the birefringence vs $c^2(E)$.

\subsection{Response of layer spacing to an electric field field applied to the Sm-A$^*$ phase}
\label{Response of Layer Spacing}

To analyze the change in layer spacing due to the application of a field, we first obtain the change in the wavevector $q=2\pi/d $. As with the orientational order we define $q_{_{E=0}}$ to be the zero-field wavevector. This is distinct from $q_0$, the zero-field wavevector in the absence of coupling between orientational order and layering. Since the wavevector only appears as $q^2$ it is convenient to define a fractional change $\Delta_{q_E}$ in $q^2$ due to the application of an electric field, i.e., $q^2=q^2_{_{E=0}}(1+\Delta_{q_E})$. In finding $\Delta_{q_E}$ we follow the same method as described in Section \ref{Response of Birefringence} and relegate the details to the Appendix. Within the Sm-A$^*$ phase, close to tricriticality, i.e. for small $\mu_{_{AC}}$, we find 
\begin{eqnarray}
\Delta_{q_E}=\frac{3|a_1|} {2K}M_{_{E=0}} \left[1+ {\cal{O}}\left( \frac{c^2(E)}{c_{_q}^2}\right) \right] c^2(E) \;,
\label{Delta q_E}
\end{eqnarray}
where, as in Ref.~\cite{SaundersTCP}, a layer contraction (as opposed to dilation) requires $a_1$ to be negative. As with $c_{_M}$, the dimensionless parameter $c_{_q}$ can be thought  of as the value of $c$ where the scaling of $\Delta_{q_E}$ with $c$ crosses over from being quadratic to quartic. We also define $c_{_q}$ in the Appendix, showing it to be $ {\cal{O}}(1)$, which for the same reasons as outlined above, allows us to neglect the quartic contribution.

Using the above equation and the relationship between layer spacing ($d$) and wavevector ($q=2\pi/d$) we next seek the contraction in the layer spacing. This contraction is equivalent to the fractional change in the layer spacing $\Delta_d = (d_{_{E=0}}-d)/d_{_{E=0}}$, where $d_{_{E=0}}$ is the zero-field value of the layer spacing in the Sm-A$^*$ phase. We find that the contraction is given by
\begin{eqnarray}
\Delta_d=\frac{3|a_1|} {4K}M_{_{E=0}}c^2(E)\;.
\label{Delta_d}
\end{eqnarray}
Since $c(E)$ is a nonlinear function of $E$ (and is not $\propto \sqrt{E}$) the above equation implies that the contraction $\Delta_d(E)$ will also be a nonlinear function of $E$, and if $c(E)$ is discontinuous, then $\Delta(E)$ will also be discontinuous. However, the above equation predicts that, like the birefringence, $\Delta_d(E)$, will scale linearly with $c^2(E)$, regardless of the nature of the transition. Thus, for small tilt angle $\theta$, which implies $c \approx \theta$, the fractional change in layer spacing scales like $\theta^2$. In addition, our theory predicts that this fractional contraction is also proportional to the size of the orientational order $M_{_{E=0}}$. Thus, de Vries systems which have unusually small orientational order will, under the application of an electric field, exhibit an unusually small layer contraction, as shown in Fig.~\ref{Delta_d_E&Delta_n_E}(b). 

Since $M_{_{E=0}}(T)$ is, as discussed in Section \ref{Response of Birefringence}, only weakly temperature dependent, the slope of the $\Delta_d(E)$ vs $c^2(E)$ should also be weakly temperature dependent. However, the slope of the absolute change in layer spacing $d_{_{E=0}}-d\equiv \Delta d (E)$  vs $c^2(E)$ should not be weakly temperature dependent. This is because  $\Delta d (E)=d(T)\Delta_d(E)$ and $d(T)$ has been shown experimentally to exhibit a noticeable monotonic increase as the Sm-A$^*$ -- Sm-C$^*$ transition is approached from above. Thus, we expect that as  temperature is lowered there should be a noticeable increase in the slope of $\Delta d (E)$ vs $c^2(E)$.

\section{Summary}
\label{Summary}

In summary, we have analyzed a generalized Landau theory for chiral smectics, one that tracks orientational, layering, tilt and biaxial order parameters as well as layer spacing. A combination of small orientational order and large layering order leads to Sm-A$^*$ -- Sm-C$^*$  transitions that are either continuous and close to tricriticality or first order. The model predicts that the change in layer spacing at the zero-field transition will be proportional to the orientational order. It also predicts that in systems having zero-field transitions that are continuous and close to tricriticality or first order, the increase in birefringence upon entry to the Sm-C$^*$ phase will be especially rapid. Thus, both the small change in layer spacing and the rapid increase in birefringence can be attributed to the system possessing a combination of small orientational order and large layering order. This is consistent with the observation that de Vries materials usually possess unusually small orientational order, which in turn means that strong layering order is required for stabilization.

The model also predicts that as a result of the zero-field Sm-A$^*$ -- Sm-C$^*$ transition being either continuous and close to tricriticality or first order, the electroclinic response of the tilt will be unusually strong. In the case of a system that has a zero-field first order Sm-A$^*$ -- Sm-C$^*$ transition, the electroclinic response tilt will exhibit a jump. Thus, as with the zero-field features of de Vries materials, our model indicates that the strong electrical response is a result of a combination of small orientational order and strong layering order. 

The equation governing the response of the tilt is completely analogous to that derived by Bahr and Heppke to describe a field induced critical point near a Sm-A$^*$ -- Sm-C$^*$  transition \cite{Bahr Heppke}. However, our derivation of the response equation from a more basic generalized Landau theory allows us to incorporate the effects of the layering and orientational orders, which we can in turn relate to the strength and nature of the tilt response. In addition, it also allows us to derive the electroclinic response of the orientational order (and thus, birefringence) and the layer spacing. We find that the change in birefringence scales quadratically with the electrically induced tilt. This means that an unusually strong tilt response implies an unusually strong response of the birefringence, as is the case in de Vries materials. The quadratic scaling is also consistent with experiment. Similarly, we find that the electrically induced change in layer spacing also scales quadratically with tilt, although the scaling is also proportional to orientational order. 

Thus, the theory predicts that a system with small orientational order and strong layering order will exhibit a combination of strong electrooptic response (in both reorientation of the optical axis and change in birefringence) and small layer change. Such a combination is technologically desirable for FLC based liquid crystal devices.

\begin{acknowledgments}

We thank Matthew Moelter for a careful reading of the manuscript. Support was provided by Research Corporation in the form of a Cottrell College Science Award.

\end{acknowledgments}
 
\appendix
\section{Field Induced Corrections to the Orientational Order and to the Layering Wavevector}
\label{Appendix}

In this appendix we provide further details of the method by which we find the fractional changes $\Delta_{M_E}$ and $\Delta_{q_E}$ to the orientational order and to the layering wavevector, respectively, due to the application of an electric field in the Sm-A$^*$ phase. This is done near a Sm-A$^*$ -- Sm-C$^*$ transition (continuous or first order) that is close to tricriticality. 
 
\subsection{Correction to the zero-field orientational order}

As discussed in Section \ref{Response of Birefringence} we are interested in finding the correction to the zero-field value of the orientational order $M_{_{E=0}}$. This zero-field value already includes the increase due to the zero-field coupling of orientational order to layering.  In the zero-field Sm-A$^*$ phase the tilt is zero and the zero-field value $M_{_{E=0}}$ was found \cite{SaundersTCP} by analyzing the part of the free energy that does not include tilt, i.e., $f_{c=0}=f_M+f_\psi+f_{M\psi}$. Specifically, we Taylor expanded $f_{c=0}(M)$ about $M_0$, the value of the orientational order in the absence of coupling to layering, i.e., the value that minimizes $f_M$. This gave
\begin{eqnarray}
f_{c=0} &\approx& f_{c=0}(M_0) + f_{M\psi}'(M_0)(M_{_{E=0}}-M_0)   \nonumber\\ 
&+& \frac{1}{2!}f_M''(M_0)(M_{_{E=0}}-M_0)^2\;,
\label{zero field M expansion}
\end{eqnarray}
where $f_{M\psi}'(M)=d f_{M\psi} /d M$ and $f_M''(M)=d^2 f_M /d M^2$. We have neglected the term $\propto f_{M\psi}''(M_0)$ which contributes terms higher order in coupling compared to $f_M''(M_0)$. Minimization of the above $f_{c=0}$ then gave $M_{_{E=0}}=M_0-\frac{f_{M\psi}'(M_0)}{f_M''(M_0)}$. 

When a field is applied to the Sm-A$^*$ phase, a tilt is induced and the tilt dependent part of the free energy becomes non-zero. Thus, to find the correction to $M_{_{E=0}}$, we Taylor expand the full free energy $f=f_{c=0}+f_c+f_{_{EC}}$ about $M_{_{E=0}}$. Doing so gives
\begin{eqnarray}
f &\approx& f_{c=0}(M_{_{E=0}}) + \left[f_c'(M_{_{E=0}})+f_{_{EC}}'(M_{_{E=0}})\right]M_{_{E=0}}\Delta_{M_E}   \nonumber\\ 
&+& \frac{1}{2!}f_{c=0}''(M_{_{E=0}})M_{_{E=0}}^2\Delta_{M_E}^2\;,
\label{nonzero field M expansion}
\end{eqnarray}
where $\Delta_{M_E}(E)=\frac{M(E)}{M_{_{E=0}}}-1$, $f_{c}'(M)=d f_{c} /d M$, $f_{_{EC}}'(M)=d f_{_{EC}} /d M$ and $f_{c=0}''(M)=d^2 f_{c=0} /d M^2$. As above, we neglect the term $\propto f_{_{EC}}''(M_{_{E=0}})$ which contributes terms higher order in coupling compared to $f_{c=0}''(M_{_{E=0}})$. Minimization of $f$ now gives
\begin{eqnarray}
\Delta_{M_E}(E)&\approx& -\frac{ \big(f_c'(M_{_{E=0}})+f_{_{EC}}'(M_{_{E=0}})\big)}{f_{c=0}''(M_{_{E=0}})M_{_{E=0}}}\;. 
\label{Delta_M expanded}
\end{eqnarray}
Keeping only terms to lowest order in coupling coefficients, $f_{c=0}''(M_{_{E=0}})\approx f_M''(M_0)\equiv \gamma_{_M}$. The dependence of $\Delta_{M_E}(E)$ on $E$ enters via the dependence of $(f_c+f_{_{EC}})$ on $E$ and $c(E)$. Since we seek to relate the correction $\Delta_{M_E}(E)$ to $c(E)$,  it is useful to express $(f_c+f_{_{EC}})$ just in terms of $c(E)$ and not $E$ explicitly. This can achieved using Eq.~(\ref{E(c)}) for $E$ in terms of $c$, giving
\begin{eqnarray}
f_c+f_{_{EC}}= -\frac{1}{2} r_c c^2(E) - \frac{1}{4} u_c c^4(E) - \frac{1}{6} v_c c^6(E)  \;. 
\label{f_c(M)+f_{_{EC}}(M)}
\end{eqnarray}
To obtain $\Delta_{M_E}(E)$, as given in Eq.~(\ref{nonzero field M expansion}), we must differentiate $f_c+f_{_{EC}}$ with respect to $M$ which enters via the coefficients $r_c(M,q^2)$, $u_c(M,q^2)$ and $v_c(M,q^2)$. These coefficients were introduced after Eq.~(\ref{f_c}), but it is convenient to present them again:
\begin{eqnarray}
r_c(M,q^2) &=&3 a(q^2) q^2 |\psi|^2 M \tau(M,q^2)\;, \nonumber\\
u_c(M,q^2) &=& 9 \mu(M,q^2) h q^4 |\psi|^2 M^2 \;,\nonumber\\
v_c(M,q^2) &=& \frac{81}{4} s q^6 |\psi|^2 M^3 \;,
\label{r_c, u_c, v_c}
\end{eqnarray}
where $a(q^2) $, $\tau(M,q^2) $ and $\mu(M,q^2) $ are given by
\begin{eqnarray}
a(q^2) &=& a_0 +a_1(q^2-q_0^2)\;, \nonumber\\
\tau(M,q^2) &=& 1-\frac{b |\psi|^2 + (g+2hq^2)M}{a(q^2)}\;, \nonumber\\
\mu(M,q^2) &=& 1-\frac{g}{2hq^2}\left(\frac{wM}{gq^2|\psi^2|}-1\right)^{-1}\;.
\label{a,tau,mu}
\end{eqnarray}
Differentiating Eq.~(\ref{f_c(M)+f_{_{EC}}(M)}) with respect to $M$, inserting the result into Eq.~(\ref{Delta_M expanded}) and keeping terms to lowest order in couplings, $t$ and  $\mu_{_{AC}}$, i.e., close to a Sm-A$^*$ -- Sm-C$^*$ transition which is close to tricriticality we find
\begin{eqnarray}
\Delta_{M_E}(E)&\approx&\frac{3m}{2\gamma_{_M}} g q^2|\psi|^ 2\left(1- \left( \frac{c(E)}{c_{_M}}\right)^2 +\left( \frac{c(E)}{c_{_{M1}}}\right)^4 \right) c^2(E)\;,
\label{Delta_M expanded1}
\end{eqnarray}
where $m=1+ \frac{2 h q^2}{g}$ is a dimensionless constant, and
\begin{eqnarray}
c_{_M}&=&\left(\frac{2g}{3hq^2}\right)^{\frac{1}{2}} \;, \nonumber\\ 
c_{_{M1}}&=&\left(\frac{4mg}{27M_{_{E=0}}sq^4}\right)^{\frac{1}{4}} \;.
\label{c_M}
\end{eqnarray}
If $g$ is of the same order as $hq^2$ then $c_{_M}$ is $ {\cal{O}}(1)$. This is not unreasonable since $g$ and $h$ are both coupling constants, which we take to be small and of the same order.  Similarly, for $M\approx M_{TC}$, which is of the same order as the coupling constants, $c_{_{M1}}\gg1$. Thus, for the small $c$ values assumed for our theory and observed experimentally, the $\left( \frac{c(E)}{c_{_M}}\right)^2$ and $\left( \frac{c(E)}{c_{_{M1}}}\right)^4$ contributions are small and the scaling of $\Delta_{M_E}(E)$ with $c(E)$ is quadratic. Note that in going from Eq.~(\ref{Delta_M expanded1}) to Eq.~(\ref{Delta q_E}) we omit the $c^6(E)$ term and replace $M$ with $M_{_{E=0}}$.

\subsection{Correction to the zero-field wavevector}

In this part of the Appendix we present details of our analysis of the fractional change, $\Delta_{q_E}(E)=\frac{q^2(E)}{q^2_{_{E=0}}}-1$, in $q^2$. As with the orientational order, we are seeking the correction to the zero-field value $q^2_{_{E=0}}$ which already includes the correction due to the zero-field coupling of orientational order to layering. The method we use to obtain $\Delta_{q_E}(E)$ is completely analogous to that used above to find $\Delta_{M_E}(E)$. Taylor expanding the free energy $f$ about $q^2_{_{E=0}}$ and minimizing with respect to $\Delta_{q_E}(E)$ we find
\begin{eqnarray}
\Delta_{q_E}(E)&\approx& -\frac{ \big(f_c'(q^2_{_{E=0}})+f_{_{EC}}'(q^2_{_{E=0}})\big)}{f_{c=0}''(q^2_{_{E=0}})q^2_{_{E=0}}}\;,
\label{Delta_q expanded}
\end{eqnarray}
where $f_{c}'(q^2)=d f_{c} /d (q^2)$, $f_{_{EC}}'(q^2)=d f_{_{EC}} /d (q^2)$ and $f_{c=0}''(q^2)=d^2 f_{c=0} /d (q^2)^2$. Keeping only terms to lowest order in coupling coefficients, $f_{c=0}''(q^2_{_{E=0}})\approx f_\psi''(q^2_0) =K|\psi|^2$. We again useEq.~(\ref{f_c(M)+f_{_{EC}}(M)}) for $f_c+f_{_{EC}}$ but now we are interested in the $q^2$ dependence of the coefficients $r_c(M,q^2)$, $u_c(M,q^2)$ and $v_c(M,q^2)$, which are given by  Eq.~(\ref{r_c, u_c, v_c}) and Eq.~(\ref{a,tau,mu}). Differentiating Eq.~(\ref{f_c(M)+f_{_{EC}}(M)}) with respect to $q^2$, inserting the result into Eq.~(\ref{Delta_q expanded}) and keeping terms to lowest order in couplings, $t$ and  $\mu_{_{AC}}$, i.e., close to a Sm-A$^*$ -- Sm-C$^*$ transition which is close to tricriticality we find
\begin{eqnarray}
\Delta_{q_E}(E)&\approx&\frac{3|a_1|} {2K} M \left(1+ \left( \frac{c(E)}{c_{_q}}\right)^2 -\left( \frac{c(E)}{c_{_{q1}}}\right)^4 \right) c^2(E)\;,
\label{Delta_q expanded1}
\end{eqnarray}
where $m=1+ \frac{2 h q^2}{g}$ is a dimensionless constant, and
\begin{eqnarray}
c_{_q}&=&\left(\frac{|a_1|}{3gq^2_{_{E=0}}M}\right)^{\frac{1}{2}} \;, \nonumber\\ 
c_{_{q1}}&=&\left(\frac{4|a_1|}{27M^2sq^4}\right)^{\frac{1}{4}} \;. 
\label{c_q}
\end{eqnarray}
For $M\approx M_{TC}$, which is of the same order as the small coupling constants, both $c_{_M}$ and $c_{_{M1}}$ are $\gg 1$. Thus, as with the correction to orientational order, for the small $c$ values assumed for our theory and observed experimentally, the $\left( \frac{c(E)}{c_{_q}}\right)^2$ and $\left( \frac{c(E)}{c_{_{q1}}}\right)^4$ contributions are small and scaling of $\Delta_{q_E}(E)$ with $c(E)$ is quadratic. Note that in going from Eq.~(\ref{Delta_q expanded1}) to Eq.~(\ref{Delta q_E}) we omit the $c^6(E)$ term and replace $q$ with $q_{_{E=0}}$.


\begin{thebibliography}{99}
\bibitem{mean field footnote} It should be noted that here, and throughout this article, exponents are calculated within mean field theory, and do not include the effects of fluctuations. For example, it is known that when fluctuation effects are included in analysis of the 3D $XY$ transition, $\theta$ scales like  $(1-\frac{T}{T_C})^\beta$, with $\beta\approx 0.35$, whereas in mean field theory $\beta = 0.5$. The use of mean field theory is justified by the fact that virtually all continuous $AC$ transitions are observed to be mean field like.
\bibitem{Notation Change} We deviate slightly from the notation used in \cite{SaundersTCP}, dropping the subscript $0$ on the $b$, $g$, $h$ and $s$ coefficients. We also use $r_n$ and $r_s$, instead of $t_n$ and $t_s$, for the coefficients of $M^2$ and $|\psi|^2$, respectively. This is done to avoid confusion with the reduced temperature $t$.
\bibitem{reduced temperature and order parameters} Of course, different materials will also have different values of $\kappa$. It is not clear how strongly these parameters will vary from material to material.
\bibitem{layering T window assumption} In a separate analysis, which we do not present here, we have found that the layering order at the Iso - Sm-A transition is indeed larger in systems with smaller orientational order. However, we have also found that for a given decrease in orientational order, the increase in layering required at the Sm-A -- Sm-C transition was larger than the increase in layering at the Iso - Sm-A transition. This would imply that a wider temperature window for the Sm-A phase will result as the orientational order is reduced.
\bibitem{tilt polarization footnote} It should be noted that in the Landau theory of \cite{Bahr Heppke}, there was a coupling between tilt and polarization. This coupling leads to an innocuous renormalization of the continuous Sm-A$^*$ -- Sm-C$^*$ transition temperature.
\end{thebibliography}
\end{document}